\documentclass[onecolumn,superscriptaddress,aps,pra,10pt,notitlepage]{revtex4-1}
\usepackage{color}
\usepackage{graphicx}
\usepackage{amsmath,amssymb}
\usepackage{ragged2e}
\usepackage{keyval}
\usepackage{braket}
\usepackage{natbib}
\usepackage{textcomp}
\usepackage[colorlinks,allcolors=blue]{hyperref}
\usepackage{bm}
\usepackage{subfigure}

\newcommand{\ii}{\ensuremath{\mathrm{i}}}

\newcommand{\K}{\ensuremath{\mathbf{K}}}
\newcommand{\kk}{\ensuremath{\mathbf{k}}}
\newcommand{\abs}[1]{\ensuremath{\left| #1 \right|}}

\begin{document}

\title{Error correction with orbital angular momentum of multiple photons propagating in a turbulent atmosphere}
\author{Jos\'e Ra\'ul \surname{Gonz\'alez Alonso}}
\email[Electronic address: ]{jrgonzal@usc.edu}
\author{Todd A. Brun}
\email[Electronic address: ]{tbrun@usc.edu}
\affiliation{Department of Physics and Astronomy, University of Southern California, Los Angeles, California, USA} 
\begin{abstract}
Orbital angular momentum of photons is an intriguing system for the storage and transmission of quantum information, but it is rapidly degraded by atmospheric turbulence.  Understanding the noise processes that affect photons is essential if we desire to protect them. In this paper we use the infinitesimal propagation equation of Roux to derive a discrete Lindblad equation, and numerically study the form of the most relevant Lindblad operators.  We find that the dominant Lindblad operators are those that shift the angular momentum by one unit.  We explore possible schemes to protect quantum information across multiple photons by concatenating a standard quantum error-correcting code with an error-detecting code for orbital angular momentum.
\end{abstract}
\maketitle

\section{Introduction}

Photons with orbital angular momentum (OAM) \cite{Allen-Orbital-1992-0,Allen-The-Orbital-1999-0,Padgett-Lights-2004-0,Yao-Orbital-2011-0} traveling through free space hold the potential to become a useful system for high bandwidth applications in both classical and quantum communications. However, to achieve this it will be necessary to protect the photons from the pernicious effects caused by a turbulent atmosphere.  Small fluctuations in the density of the air lead to random changes in the index of refraction across the wave front of a propagating photon; these phase shifts distort the state of the photon, causing shifts in its angular momentum and radial degrees of freedom.  This distortion represents a strong source of noise for quantum communication that requires robust error correction or suppression \cite{Gonzalez-Alonso-Protecting-2013-0}.

In the quantum case, understanding this noise requires precise modeling of the decoherence effects of such an atmosphere as well as an application of quantum error detection or correction schemes to protect the quantum information encoded in each photon. To this end, we will be using the infinitesimal propagation equation of Roux \cite{Roux-Infinitesimal-propagation-2011-0,Roux-Erratum:-2013-0} to model the noise process that a photon undergoes while traveling through the air, and will derive a Lindblad representation of this process with a discrete set of Lindblad operators.

While in principle the orbital angular momentum states offer an infinite-dimensional Hilbert space in which to store quantum information, in practice only a finite number of such states may be practically prepared or manipulated.  Larger quantum states, therefore, will have to be encoded across multiple photons---particularly likely if the quantum information is encoded in a quantum error-correcting code (QECC).  However, multiple photons traveling together with a time separation that is less than the characteristic time of the turbulence process will undergo the same random phase shifts, as they propagate through the same volume of air with the same density fluctuations.  This means that the noise process for such a train of photons will be symmetric under permutations of the photons.  From the Lindblad representation for a single photon, we can therefore find the representation for a train of closely spaced photons.

We then briefly discuss possible methods for error correction with a codeword encoded in the angular orbital momenta of a train of photons, whether it is possible to exploit this exchange symmetry to boost the resistance to noise, and the possibility of combining error correction with adaptive optics.  We end by
giving a few numerical examples of our scheme and briefly discussing our results.

\section{Infinitesimal Propagation Equation} 

In \cite{Roux-Infinitesimal-propagation-2011-0,Roux-Erratum:-2013-0} the so-called infinitesimal propagation equation (IPE) for an OAM entangled biphoton travelling in a turbulent atmosphere is presented. The formalism is then used in \cite{Roux-Lindblad-2014-0} to derive a Lindblad-like equation for the evolution of a single photon with a continuous spectrum of Lindblad operators. However, unlike the work in \cite{Roux-Lindblad-2014-0}, we would like to obtain a discrete set of Lindblad operators and investigate their effects in order to gain insight into the error process an OAM photon experiences in a turbulent atmosphere and (hopefully) how to correct it.

We start with the IPE for a single OAM photon using the notation and results of Roux from \cite{Roux-Infinitesimal-propagation-2011-0,Roux-Erratum:-2013-0}:
\begin{equation}
\partial_z \rho_{mn}(z) = -\ii S_{mu}(z)\rho_{un}(z) + \ii \rho_{mv}(z)S_{vn}(z) + L_{mnuv}(z)\rho_{uv}(z) - L_T \rho_{mn}(z).
\label{eq:IPE_OAM}
\end{equation}
Here, $z$ represents the propagation distance of the photon along the beam path; this distance plays the role of time in a usual master equation.

In Eq.~\eqref{eq:IPE_OAM}, the indices $m,n,u,$ and $v$ each represent a collective index for both the radial and orbital degrees of freedom of the OAM state represented by the density matrix $\rho$.  Repeated indices imply a summation. The operator that represents the free-space propagation is
\begin{align}\label{eq:free_space_int}
S_{x,y}(z) = \frac{\ii}{2k}\int\abs{\K}^2 G_x(\K,z)G_y^*(\K,z)\frac{\mathrm{d}^2 K}{4\pi^2} .
\end{align}
The dissipative part of the evolution is given by
\begin{align}\label{eq:diss_int}
L_{mnuv}(z) = k^2 \int \Phi_1(\K) W_{m,u}(\K,z)W_{n,v}^*(\K,z)\frac{\mathrm{d}^2 K}{4\pi^2} .
\end{align}
Finally, there is a divergent dissipative term that is given by
\begin{align}\label{eq:div_diss_int}
L_T = k^2 \int \Phi_1(\K) \frac{\mathrm{d}^2 K}{4\pi^2} .
\end{align}

In Eqs.~(\ref{eq:free_space_int}--\eqref{eq:div_diss_int}), the vector $\K = (k_x, k_y)$ is the 
two dimensional projection of the propagation vector $\kk=(k_x,k_y,k_z)$. The function $G_x(\K,z)$ is the two-dimensional momentum space wave function of the OAM basis. Therefore, because of the orthogonality of the OAM basis, the momentum space wave functions satisfy
\begin{equation}
\int G_x(\K,z) G_y^*(\K,z) \frac{\mathrm{d}^2 K}{4\pi^2} = \delta_{x,y}.
\end{equation}
The convolution of the momentum space wave functions is denoted by $W_{x,y}(\K,z)$:
\begin{align}
W_{x,y}(\K,z) = \int G_{x} (\K_1,z) G_{y}^* (\K_1 - \K_,z) \frac{\mathrm{d}^2 K_1}{4\pi^2}.
\end{align}

In the IPE formalism it is possible in principle to use any power spectral density $\Phi_1(\K)$ for the turbulence model. For the current work, if we ignore the effect of the inner scale of the turbulence, we can start with the von~Karman power spectral density \cite{Andrews-Laser-2005-0} with the Fourier convention from \cite{Roux-Infinitesimal-propagation-2011-0,Roux-Erratum:-2013-0}:
\begin{equation}
\label{eq:turb_pow_sp_dens}
\Phi_1(\K) =
\frac{20 \pi^2 C_n^2}{9 \Gamma\left(\frac{1}{3}\right) \left(\abs{\K}^2 + \kappa_0^2\right)^{11/6}}
\approx \frac{8.186994\, C_n^2}{\left(\abs{\K}^2+\kappa_0^2\right)^{11/6}} .
\end{equation}
In Eq. \eqref{eq:turb_pow_sp_dens}, $\kappa_0^2$ is used for the outer scale of the turbulence. In our calculations, it allow us to regularize the integrals in Eqs.~\eqref{eq:diss_int} and \eqref{eq:div_diss_int} so that we can eventually take the large outer scale limit $\kappa_0\rightarrow 0$. This is an important case because it allows us to study the Kolmogorov model of turbulence.

\subsection{Generating Functions for the Integrals}

To analyze the effects of turbulence via the IPE, we must write the integrals in Eqs.~(\ref{eq:free_space_int}--\ref{eq:div_diss_int}) in a form that is more amenable to calculations. In \cite{Roux-Infinitesimal-2011-0}, a generating function for the Laguerre-Gauss modes is used to obtain generating functions for the different integrals in Eqs.~(\ref{eq:free_space_int}--\ref{eq:div_diss_int}). In what follows, we briefly review these results.

The generating function for the Laguerre-Gauss modes used in \cite{Roux-Infinitesimal-propagation-2011-0,Roux-Erratum:-2013-0} is
\begin{eqnarray}
G & = & \sum_{n,m=0}^{\infty} \frac{1}{m!} L_n^m\left(\frac{2(u^2+v^2)}{1+t^2}\right) \left[\frac{w(1+\ii t)}{1-\ii t}\right]^n
\frac{\left[(u+\ii v)p + (u-\ii v)q\right]^m}{(1-\ii t)^{1+m}} \nonumber \\
& = & \frac{1}{\Omega(t,w)} \exp \left[ \frac{(u+\ii v)p}{\Omega(t,w)} + \frac{(u-\ii v)q}{\Omega(t,w)} 
- \frac{(1+w) (u^2+v^2)}{\Omega(t,w)} \right] ,
\end{eqnarray}
where $\Omega(t,w) = 1-w-\ii t-\ii wt$. The normalized coordinates $u,v$, and $t$ are given by $u=x/\omega_0$, $v=y/\omega_0$ and $t=z\lambda/\pi\omega_0^2$. In these expressions $\omega_0$ is the initial beam waist and $\lambda$ is the wavelength of the beam of light. The parameters $p$, $q$, and $w$ generate the Laguerre-Gauss modes by taking derivatives of the generating function according to the following rules:
\begin{equation}
\label{eq:gen_fun_rule}
M^{LG}_{r,l}(u,v,t) = \left\{ \begin{array}{lcc}
{\cal N} \left[ \frac{1}{r!} \partial_w^r \partial_p^{|l|} G \right]_{w,p,q=0} & {\rm for} & l>0 \\
{\cal N} \left[ \frac{1}{r!} \partial_w^r  G \right]_{w,p,q=0} & {\rm for} & l=0 \\
{\cal N} \left[ \frac{1}{r!} \partial_w^r \partial_q^{|l|} G \right]_{w,p,q=0} & {\rm for} & l<0 , \\
\end{array} \right.
\end{equation}
where $r$ and $l$ are the radial and azimuthal indices, respectively, and ${\cal N}$ is the normalization constant:
\begin{align}
{\cal N} = \left[ \frac{r!2^{|l|+1}}{\pi (r+|l|)!} \right]^{1/2}.
\end{align}

The rule in Eq.~\eqref{eq:gen_fun_rule} also applies when using the generating functions of the integrals for the different terms in the IPE. However, for this we will need the Fourier transform of the generating function, which is
\begin{equation}
\label{eq:fourier_gen_fun}
{\cal F} \{G\} =  \frac{\pi}{1+w}\exp \left[ \frac{\ii\pi (a+\ii b)p}{1+w} + \frac{\ii\pi (a-\ii b)q}{1+w} 
- \frac{\pi^2 (a^2+b^2)\Omega(t,w)}{1+w} \right].
\end{equation}
Here, $a$ and $b$ are related to the wave vector components by $k_x = {2\pi a}/{\omega_0}$ and $k_y = {2\pi b}/{\omega_0}$.

\subsubsection{Free-Space Propagation Term}

Using the generating function in \eqref{eq:fourier_gen_fun}, and then exchanging the order of the integral and the derivatives, we can obtain a generating function for Eq.~\eqref{eq:free_space_int}. Furthermore, from the form of this function, we find that the azimuthal indices involved must be equal, and that the radial indices can differ at most by one.  In other words, the result of the integral in \eqref{eq:free_space_int} simplifies to
\begin{align}\label{eq:gen_fun_free}
S_{m,n}(z) &= 
\begin{cases}
    \frac{\ii (1+\abs{l} + 2r)}{2z_R} , & l_m=l_n=l,r_m=r_n=r , \\
    \frac{\ii (1+\abs{l} + r)^\frac{1}{2}(1+r)^\frac{1}{2}}{2z_R} , & 
        \left\{\begin{array}{l}
            l_m=l_n=l , \\
            \abs{r_m-r_n}=1 , \\
            r=\frac{r_m+r_n-1}{2} ,
        \end{array}\right.\\
    0, & \text{otherwise},
\end{cases}
\end{align}
where the azimuthal indices are indicated by $l_m=l_n=l$ and the radial indices are indicated by $r_m$ and $r_n$.

\subsubsection{Divergent and Dissipative Terms}

With the spectral density we chose in \eqref{eq:turb_pow_sp_dens}, the divergent dissipative term is
\begin{equation}
L_T = \frac{8 \pi^3 C_n^2}{3 \kappa_0^{10/6} \lambda^2 \Gamma\left(\frac{1}{3}\right)}
    \approx \frac{30.86424\, C_n^2}{\kappa_0^{10/6} \lambda^2}.
\label{eq:divergent_term_no_integral}
\end{equation}

However, to generate the function for the dissipative and divergent terms, we now need to use a generating function for the radial indices of the modal correlation functions \cite{Roux-Infinitesimal-propagation-2011-0,Roux-Erratum:-2013-0}:
\begin{eqnarray}
W_{rG}(K,\phi,z) & = & \frac{\exp(-X)\exp[\ii(l_m-l_n)\phi] \overline{E}_{n}^{|l_n|} E_{m}^{|l_m|}}{(1-w_m w_n)} \left[ \frac{r_n!}{(|l_n|+r_n)!}\right]^{1/2} \left[\frac{r_m!}{(|l_m|+r_m)!}\right]^{1/2} \nonumber \\
 & & \times \sum_{s=0}^{M(l_m,l_n)} \frac{|l_m|!|l_n|! (-X)^{-s}}{(|l_m|-s)! (|l_n|-s)! s!} ,
\label{somw}
\end{eqnarray}
where $l_m$ and $l_n$ are the azimuthal indices and $r_m$ and $r_n$ are their associated radial indices.
Moreover \cite{Roux-Infinitesimal-propagation-2011-0,Roux-Erratum:-2013-0},
\begin{align}
M(l_m,l_n) & =   \frac{1}{2} \left( |l_m|+|l_n|-|l_m-l_n| \right) \label{parmm} , \\
X & = \frac{K^2 \zeta_m \zeta_n^* \eta^2}{8\pi^2 (1-w_m w_n)} \label{parmx} , \\
E_m & =  \frac{\ii \sqrt{2} K \zeta_m \eta}{4\pi(1-w_m w_n)} \label{parme} , \\
\overline{E}_n & =  \frac{\ii \sqrt{2} K \zeta_n^* \eta}{4\pi(1-w_m w_n)} , \label{parmes},
\end{align}
where $\zeta_a = z_R -\ii z - w_a (z_R + \ii z)$, 
$\eta = \lambda/\ \omega_0$, and we are using polar momentum space coordinates:
\[
k_x + \ii k_y = K\exp(\ii\phi) ,
\]

Once again, if we first perform the integral of each term in the sum in each of the generating functions for the convolution terms of Eq.~\eqref{eq:diss_int}, we can get a generating function for the radial part of for $L_{mnuv}(z)$ in \eqref{eq:diss_int}. Since we are interested in the Kolmogorov model for turbulence, we take the limit $\kappa_0\rightarrow 0$ in our expressions.  This will produce terms that have divergences when $m=u$ and $n=v$, but fortunately, these divergences are canceled by those in $L_T$.  Therefore, after subtracting the divergent term $L_T$ for the appropriate combinations of indices $m,n,u,$ and $v$, and  taking the limit $\kappa_0\rightarrow 0$, we get a generating function for the radial part of Eq.~\eqref{eq:diss_int} expressed as two sums, one for each of the convolutions in the integral.  That is:
\begin{align}
\begin{split}
L^{rG}_{mnuv} = &
\sum_{s=0}^{M(l_m,l_u)}\sum_{s'=0}^{M(l_n,l_v)}
\frac{(-1)^{s+s'}}{D(l_m,r_m,l_n,r_n,l_u,r_u,l_v,r_v,s,s')}
\Bigg[
A(l_m,r_m,l_n,r_n,l_u,r_u,l_v,r_v)\Bigg.\\
 &\times \Omega(t,w_m)^{\left| l_m\right| -s} (\Omega^*(t,w_n))^{\left| l_n\right| - s'}
 (\Omega^*(t,w_u))^{\left| l_u\right| -s} \Omega(t,w_v)^{\left| l_v\right| - s'}\\
 &\times (1-w_m w_u)^{-\left| l_m\right| -\left| l_u\right| + s - 1}
 (1-w_n w_v)^{-\left| l_n\right| -\left| l_v\right| + s' - 1}
 \omega_0^{\left| l_m\right| +\left| l_n\right| +\left| l_u\right| +\left| l_v\right| -2 (s+s')}\\
 &\times\Bigg.\Gamma \left(\frac{1}{2} (\left| l_m\right| + \left| l_n\right| + 
 \left| l_u\right| + \left| l_v\right|) - (s+s') -\frac{5}{6})\right)
B(w_m,w_n,w_u,w_v,t,\omega_0)^{P(l_m,l_n,l_u,l_v,s,s')})\Bigg] ,
\end{split}
\end{align}
where
\begin{align}
&A(l_m,r_m,l_n,r_n,l_u,r_u,l_v,r_v) = 
\ii^{\left| l_m\right| +\left| l_u\right| - \left| l_n\right| - \left| l_v\right| }
5 \pi ^3 C_n^2 \sqrt{r_m!\,r_n!\,r_u!\,r_v!}
\left| l_m\right| ! \left| l_n\right| ! \left| l_u\right| ! \left| l_v\right| ! ,
\end{align}
\begin{align}
\begin{split}
B(w_m,w_n,w_u,w_v,t,\omega_0,s,s') & = -\frac{\omega_0^2}{(w_m w_u-1) (w_n w_v-1)} \\
&\times \left( t^2 (w_m (w_n (2 w_u w_v+w_u+w_v)+w_u w_v-1) + w_n w_u w_v-w_n-w_u-w_v-2)\right.\\
 &+2 \ii t (w_m (w_n w_u-w_n w_v-w_u w_v+1)+w_n w_u w_v-w_n-w_u+w_v)\\
 &\left.+w (2 w_n w_u w_v-w_n w_u-w_n w_v-w_u w_v+1)-w_n w_u w_v+w_n+w_u+w_v-2\right) ,
 \end{split}
 \end{align}
 \begin{align}
 P(l_m,l_n,l_u,l_v,s,s') = &
 -\frac{\left| l_m \right| - \left| l_n\right| - \left| l_u\right| - \left| l_v\right|}{2} + s + s'+ \frac{5}{6} ,
 \end{align}
 \begin{align}
 \begin{split}
 D(l_m,r_m,l_n,r_n,l_u,r_u,l_v,r_v,s,s') &= 9 \sqrt{2} \lambda ^2 
 \Gamma \left(\frac{1}{3}\right) s! s'! \sqrt{(\left| l_m\right| + r_m)! (\left| l_n\right| +
 r_n)! (\left| l_u\right| + r_u)! (\left| l_v\right| +r_v)!}\\
 &\times (\left| l_m\right| - s)! (\left| l_n\right| - s')! (\left| l_u\right| -s)! (\left| l_v\right| - s')! .
 \end{split}
\end{align}
While these expressions appear very complicated, they are not particularly difficult to handle with an appropriate computer program.

\section{Lindblad Equation}\label{sec:lindblad_equation}

Our goal in using generating functions for each of the terms in Eq.~\eqref{eq:IPE_OAM} is to rewrite it explicitly in Lindblad form using a discrete set of Lindblad operators (unlike the continuous Lindblad operators in \cite{Roux-Lindblad-2014-0}). This discrete set of Lindblad operators may then yield a better understanding of the effect of atmospheric turbulence on OAM as an error process with a set of dominant error types (represented by the dominant Lindblad operators).  This, in turn, should be useful in designing QECCs to protect quantum information against these dominant errors.

Additionally, we can also use these Lindblad operators to build a simplified model for the noise process on multiple photons when the time separation between photons is less than the characteristic time of the turbulence, and to design suitable encodings across multiple photons.

To achieve this, we want to rewrite Eq.~\eqref{eq:IPE_OAM} in the superoperator formalism in the following manner:
\begin{align}
\partial_z \mathrm{col}(\rho) &= \mathrm{col}(\mathbb{C}(\rho) + \mathbb{D}(\rho)) ,
\label{eq:superoperator_form}
\end{align}
where the notation ``$\mathrm{col}(\rho)$'' means to write the $N\times N$ matrix $\rho$ as an $N^2\times 1$ column 
vector by stacking the columns of $\rho$ on top of each other.  A linear map on $\rho$ then becomes an 
$N^2\times N^2$ matrix multiplying the vector $\mathrm{col}(\rho)$.

In Eq.~\eqref{eq:superoperator_form}, $\mathbb{C}(\rho) = -\ii\left[\mathbf{H},\rho\right]$ represents the coherent evolution part of the Lindblad equation, while $\mathbb{D}(\rho)=\sum_k \mathbf{L}_k\rho \mathbf{L}_k^ \dagger - \frac{1}{2}\mathbf{L}_k^\dagger\mathbf{L}_k\rho - \frac{1}{2}\rho \mathbf{L_k^\dagger}\mathbf{L_k}$ represents the decoherent part of the evolution.  Using the identity \cite{Petersen-The-Matrix-2012-0}
\[
\mathrm{col}(\mathbf{AXB})=\mathbf{B}^T\otimes\mathbf{A}\, \mathrm{col}(\mathbf{X}) ,
\]
we can rewrite these superoperators using their respective matrix representations $\mathcal{C}$ and $\mathcal{D}$:
\begin{align}\label{eq:coherent_super_matrix}
\mathcal{C} = -\ii \left( \mathbb{I}\otimes \mathbf{H}  - \mathbf{H}\otimes\mathbb{I}\right) ,
\end{align}
and
\begin{align}\label{eq:decoherent_super_matrix}
\mathcal{D}=\sum_k
\overline{\mathbf{L}}_k\otimes \mathbf{L}_k -
\frac{1}{2}\mathbb{I}\otimes\mathbf{L}_k^\dagger\mathbf{L}_k -
\frac{1}{2} \overline{\mathbf{L}_k^\dagger\mathbf{L}_k}\otimes\mathbb{I} ,
\end{align}
so Eq.~\eqref{eq:superoperator_form} becomes a simple linear equation:
\[
\partial_z \mathrm{col}(\rho) = \left( \mathcal{C} + \mathcal{D} \right) \mathrm{col}(\rho) .
\]

\subsection{Obtaining Lindblad Operators from a Superoperator Representation}

By an appropriate grouping of the indices in Eq.~\eqref{eq:IPE_OAM}, one can numerically find $N^2\times N^2$ approximations to the matrices $\mathcal{C}$ and $\mathcal{D}$ for a given range of collective (i.e., both azimuthal and radial) OAM indices.  (Since the Hilbert spaces of both the azimuthal and radial degrees of freedom are in principle infinite dimensional, we approximate by truncating to a finite dimensional subspace.)

As we have done before, let us denote by $u,v$ the collective OAM indices used to describe the quantum state before the effects of the turbulence, while denoting by $m,n$ the collective OAM indices used to describe the state after the effects of turbulence.  The initial density matrix $\rho$ can be written
\[
\rho = \sum_{u,v} \rho_{(u,v)} \ket{u}\bra{v} .
\]
These outer products $\ket{u}\bra{v}$ are mapped into basis vectors $\ket{u,v}$ for $\rho$:
\[
\mathrm{col}(\rho) = \sum_{u,v} \rho_{(u,v)} \ket{u,v} .
\]
So the elements of the column vector $\mathrm{col}(\rho)$ are labeled by a pair of collective indices $u,v$, and the elements of the matrices $\mathcal{C}$ and $\mathcal{D}$ are labeled by two pairs of indices $(m,n)$ and $(u,v)$.

 We will use azimuthal and radial indices such that $-L\le l_m,l_n,l_u,l_v\le L$ and $0\le r_m,r_n,r_u,r_v\le L$ with $L>0$. We choose $L$ such that both pairs of collective indices $(u,v)$ and $(m,n)$ are less than or equal to the cutoff $N^2$.  The cutoff of the indices might result in ``leakage'' errors of the state out of the finite dimensional subspace.  In principle, the effects of this leakage could be described by adding an extra term to Eq.~\eqref{eq:superoperator_form} involving an anti-commutator with an (unknown) operator; but we will ignore that term in the rest of this work, on the assumption that our cutoff is sufficiently higher than the indices of the initial state that leakage errors are negligible.

Under all of the above assumptions, we can rewrite the  part of \eqref{eq:IPE_OAM} that represents the free-space propagation  of the beam to be included in $\mathcal{C}$, as
\begin{align}
\left(\mathcal{C}\right)_{(m,n),(u,v)} = S_{u,m}\delta_{v,n} - S_{v,n}\delta_{m,u}.
\end{align}
For the decoherent superoperator matrix we have
\begin{align}
\left(\mathcal{D}\right)_{(m,n),(u,v)} = \frac{1}{r_m!r_n!r_u!r_v!}
\left.\partial^{r_m,r_n,r_u,r_v}_{w_m,w_n,w_u,w_v} L^{rG}_{m,n,u,v}\right|_{w_m=w_n=w_u=w_v=0} .
\end{align}
From this, and comparing with \eqref{eq:coherent_super_matrix}, and \eqref{eq:decoherent_super_matrix} we can extract a form for the Lindblad operators using the procedure described in \cite{Havel-Robust-2003-0}. Essentially, this requires us to obtain the eigenvectors $\{v_k\}$ and eigenvalues $\{\lambda_k\}$ of
\begin{align}
\tilde{\mathcal{D}} = \mathcal{P}^I \mathrm{Choi}(\mathcal{D}) \mathcal{P}^I,
\end{align}
where 
$\mathcal{P}^I = \mathbb{I}\otimes\mathbb{I} - 1/N \mathrm{col}(\mathbb{I})\mathrm{col}(\mathbb{I})^\dagger$, and $\mathrm{Choi}(A)$ is the Choi matrix of $A$ \cite{Choi-Completely-1975-0}.  A Lindblad operator $\mathbf{L}_k$ is constructed by taking the square root of the eigenvalue and reorganizing the eigenvectors to form a matrix of the appropriate size:
\[
\sqrt{\lambda_k} v_k = \mathrm{col}(\mathbf{L}_k) .
\]
With this procedure, we can extract the most important Lindblad operators, as measured by the size of the eigenvalues $\lambda_k$ used to construct them.  As we will see below, if we order the eigenvalues $|\lambda_1| \ge |\lambda_2| \ge \cdots \ge |\lambda_{N^2}|$ the magnitudes of the eigenvalues $|\lambda_k|$ fall off rapidly with $k$.  So the first few Lindblad operators dominate the error process.

\subsection{Multiple closely-spaced photons}

As a photon propagates through a thin slice of turbulent air, its state is transformed by the random fluctuations in the density (and hence in the index of refraction) across the wave front.  If we knew the precise values of the density fluctuations, we could describe the evolution of the state by some unitary transformation
\begin{equation}
\rho(z) \rightarrow \rho(z+\Delta z) = \exp\left(-i\varepsilon \mathbf{H}\right) \rho(z) \exp\left(i\varepsilon \mathbf{H}\right)
\approx -i\varepsilon [\mathbf{H}, \rho(z)] + \varepsilon^2\left( \mathbf{H}\rho(z)\mathbf{H} - (1/2)\{\mathbf{H}^2,\rho(z)\} \right) ,
\label{eq:turbulence_hamiltonian}
\end{equation}
where $\varepsilon$ is a small parameter and $\mathbf{H}$ is a random Hamiltonian that describes the unitary evolution of the photon state across the thin slice of air from $z$ to $z+\Delta z$.

Of course, in practice we do not know $\mathbf{H}$, and so we take an ensemble average of Eq.~\eqref{eq:turbulence_hamiltonian} over all realizations of $\mathbf{H}$.  Assuming that the ensemble average of $\mathbf{H}$ vanishes (that is, mean zero noise), the linear term in \eqref{eq:turbulence_hamiltonian} goes away, but the quadratic terms do not, and this average transforms Eq.~\eqref{eq:turbulence_hamiltonian} into a Lindblad equation.  We see by inspection that the Lindblad operators $\{\mathbf{L}_k\}$ must be linear combinations of these possible Hamiltonians $\mathbf{H}$.  (If the mean is small but not exactly zero, it is possible for there also to be a Hamiltonian term in the Lindblad equation.)

Now suppose that a sequence of $n$ photons all cross the slice of air from $z$ to $z+\Delta z$ within a time that is short compared to the characteristic time of the turbulence.  Then the state of each photon experiences exactly the same unitary transformation given in Eq.~\eqref{eq:turbulence_hamiltonian}, which is the same as having a collective Hamiltonian $\mathbf{\tilde{H}}$ acting on all the photons of the form
\begin{equation}
\mathbf{\tilde{H}} = \sum_{i=1}^{n} \left(\bigotimes_{j=1}^{i-1} \mathbb{I}\right)\otimes \mathbf{H} \otimes
    \left(\bigotimes_{j=i+1}^{n} \mathbb{I} \right).
\end{equation}
The evolution of the $n$-photon state then looks like
\begin{equation}
\rho(z) \rightarrow \rho(z+\Delta z) = \exp\left(-i\varepsilon \mathbf{\tilde{H}}\right) \rho(z) \exp\left(i\varepsilon \mathbf{\tilde{H}}\right)
\approx -i\varepsilon [\mathbf{\tilde{H}}, \rho(z)] + \varepsilon^2\left( \mathbf{\tilde{H}}\rho(z)\mathbf{\tilde{H}} - (1/2)\{\mathbf{\tilde{H}}^2,\rho(z)\} \right) .
\label{eq:turbulence_hamiltonian_n}
\end{equation}

Taking the ensemble average of the $n$-photon equation \eqref{eq:turbulence_hamiltonian_n} will again yield a Lindblad equation.  Furthermore, if $\{\mathbf{L}_k\}$ for $k=1,\ldots,N^2$ is the set of Lindblad operators for the one-photon case, we can approximate the Lindblad operators for the $n$-photon case by
\begin{align}
    \mathbf{\tilde{L}}_k = \sum_{i=1}^{n} \left(\bigotimes_{j=1}^{i-1} \mathbb{I}\right)\otimes \mathbf{L}_k \otimes
    \left(\bigotimes_{j=i+1}^{n} \mathbb{I} \right).
\end{align}
We will discuss the properties of the Lindblad operators for the one-  and $n$-photon cases in the following sections.

\begin{figure}[htbp]
\includegraphics[width=0.475\columnwidth]{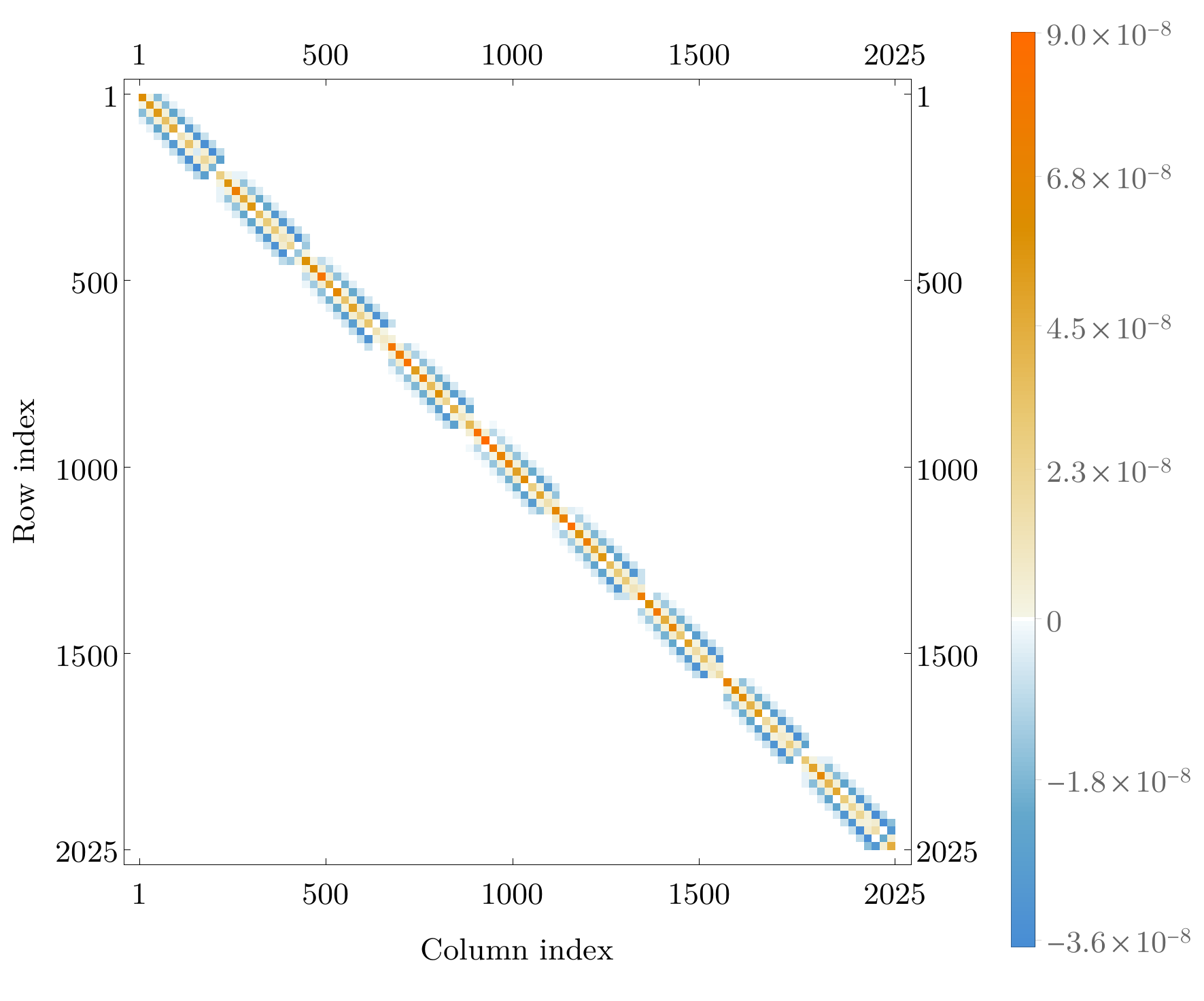}
\caption{Example of the operator $\mathcal{C}$ when $\lambda=1.0\times10^{-6}$ m, $\omega_0=0.01$ m, $C_n^2=1.0\times 10 ^{-14}\;m^{-2/3}$, $t_z=100$, and $L=4$.}
\label{fig:PlotImHmat1}
\end{figure}

\subsection{Numerical Examples of Lindblad Operators}

\begin{figure}[htbp]
\subfigure[~Real part of $\mathcal{D}$]{
\includegraphics[width=0.475\columnwidth]{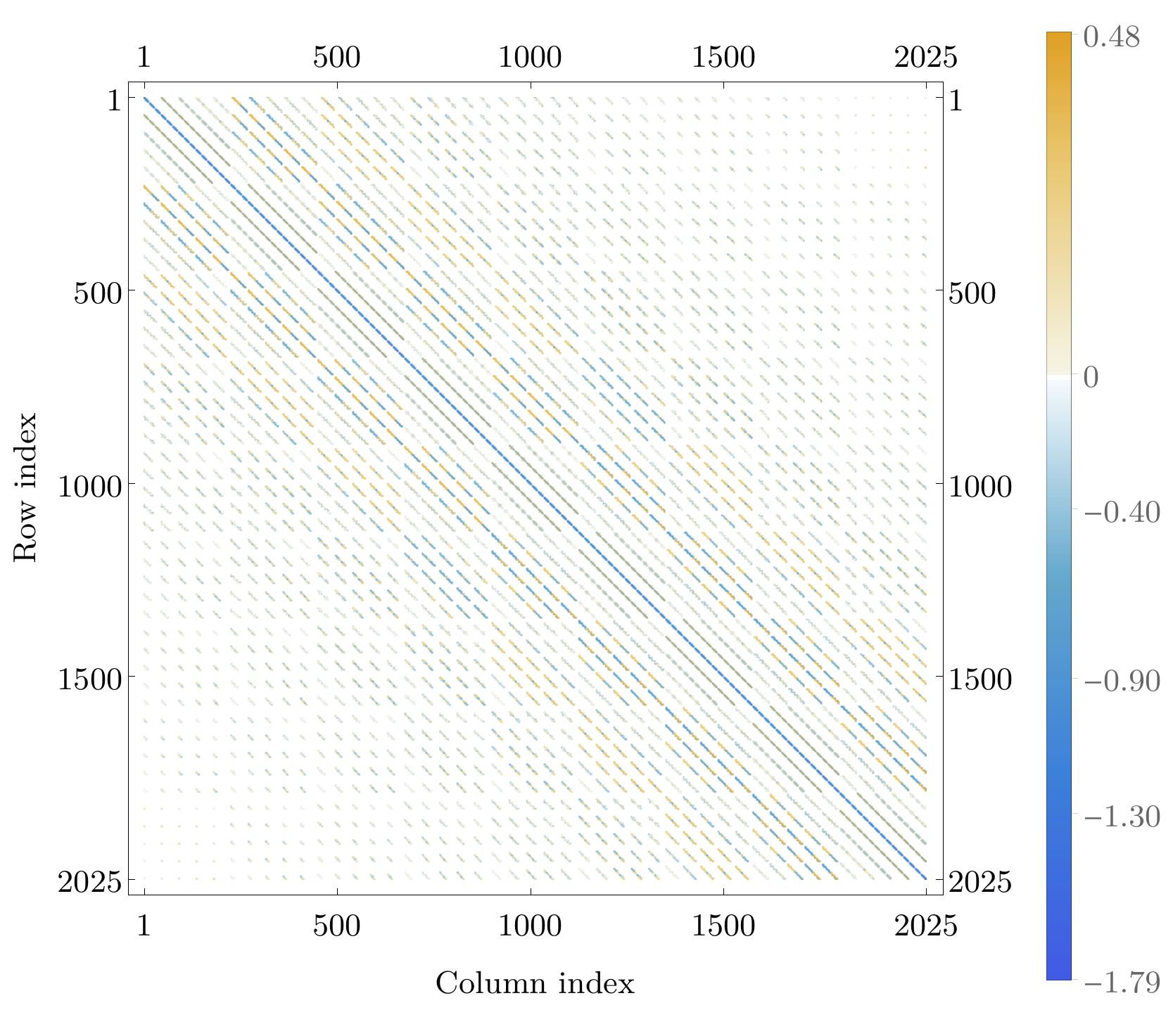}
\label{fig:PlotReGmat5}}
\subfigure[~Imaginary part of $\mathcal{D}$]{
\includegraphics[width=0.475\columnwidth]{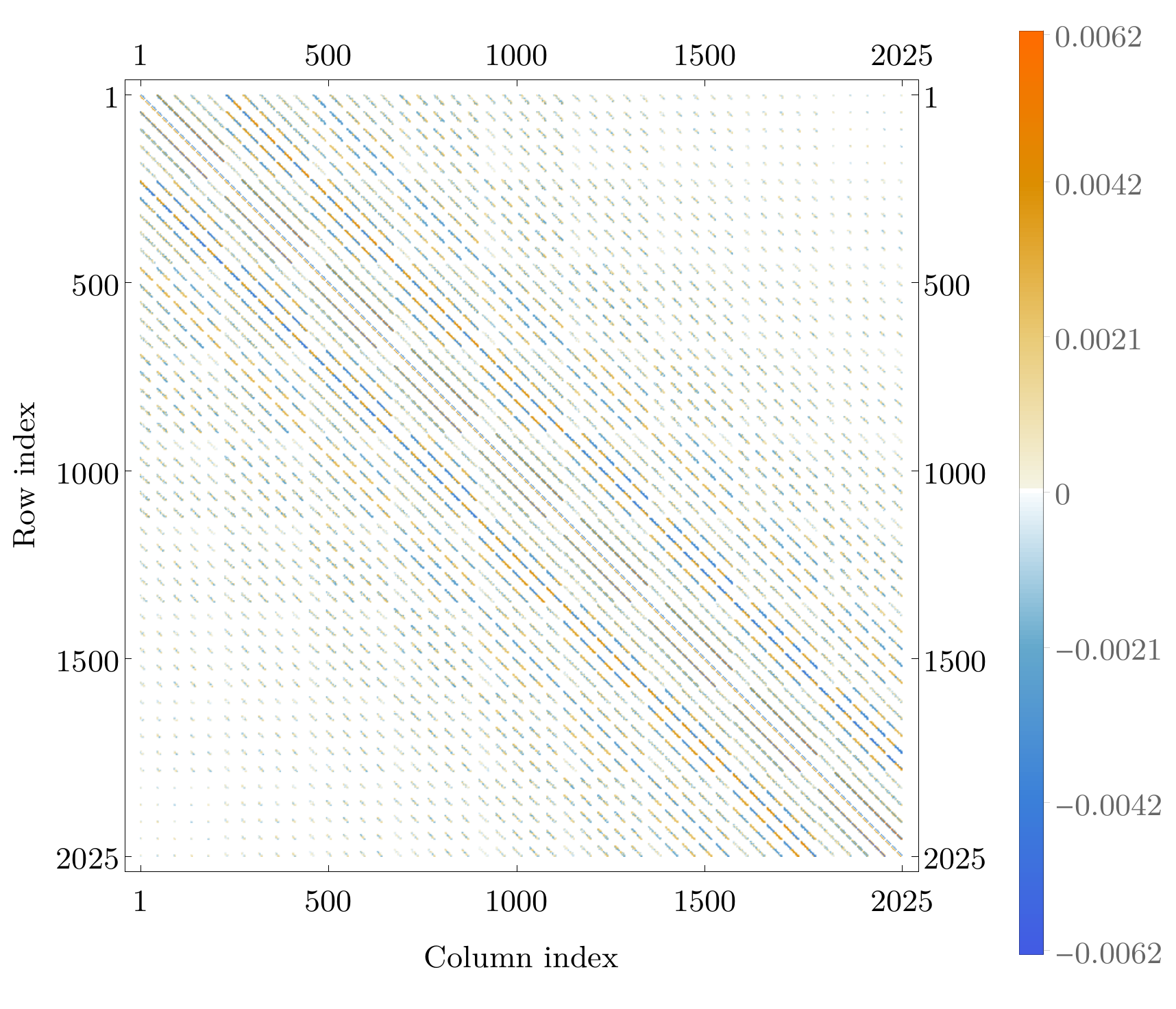}
\label{fig:PlotImGmat5}}
\caption{Example of the operator $\mathcal{D}$ when $\lambda=1.0\times10^{-6}$ m, $\omega_0=0.01$ m, $C_n^2=1.0\times 10 ^{-14}\;m^{-2/3}$, $t_z=100$, and $L=4$.}
\label{fig:PlotGmat5}
\end{figure}

To get a better idea of how the matrix representation of the superoperators $\mathcal{C}$ and $\mathcal{D}$ look, we have calculated some examples numerically.  In Fig.~\ref{fig:PlotImHmat1}, we show how the distribution of the non-zero elements in the matrix for the coherent part of the evolution in the IPE looks like. As expected from Eqs.~\eqref{eq:gen_fun_free} and \eqref{eq:coherent_super_matrix}, we obtain a purely imaginary sparse matrix where all of the non-zero elements are concentrated around the diagonal.

\begin{figure}[htbp]
\includegraphics[width=0.475\columnwidth]{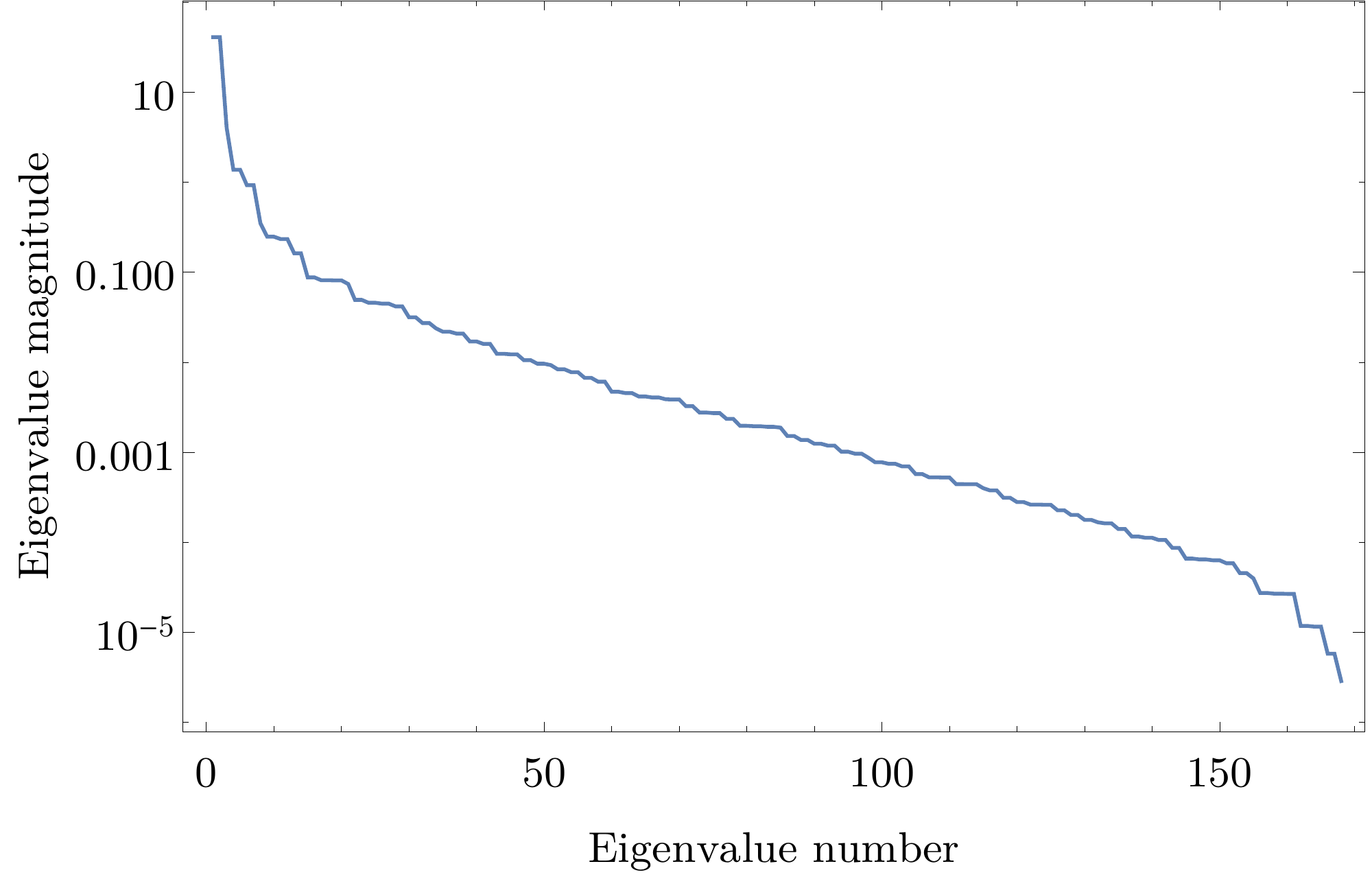}
\caption{Magnitudes of the eigenvalues of the operator $\mathcal{\tilde{D}}$ when $\lambda=1.0\times10^{-6}$ m, $\omega_0=0.01$ m, $C_n^2=1.0\times 10 ^{-14}\;m^{-2/3}$, $t_z=100$, and $L=4$.}
\label{fig:EigvalMyProj5}
\end{figure}

On the other hand, as can be seen in Fig.~\ref{fig:PlotGmat5} the decoherent part of the IPE evolution involves elements that in general are complex numbers. While the magnitude of these elements changes with the propagation distance, the structure of the superoperator matrix remains the same. Also, we can see that the matrix elements representing transitions for either the azimuthal or radial degrees of freedom become smaller in magnitude as the size of the shifts in these numbers becomes larger.

\begin{figure}[htbp]
\subfigure[~Real part of $\mathbf{L}_1$]{
\includegraphics[width=0.475\columnwidth]{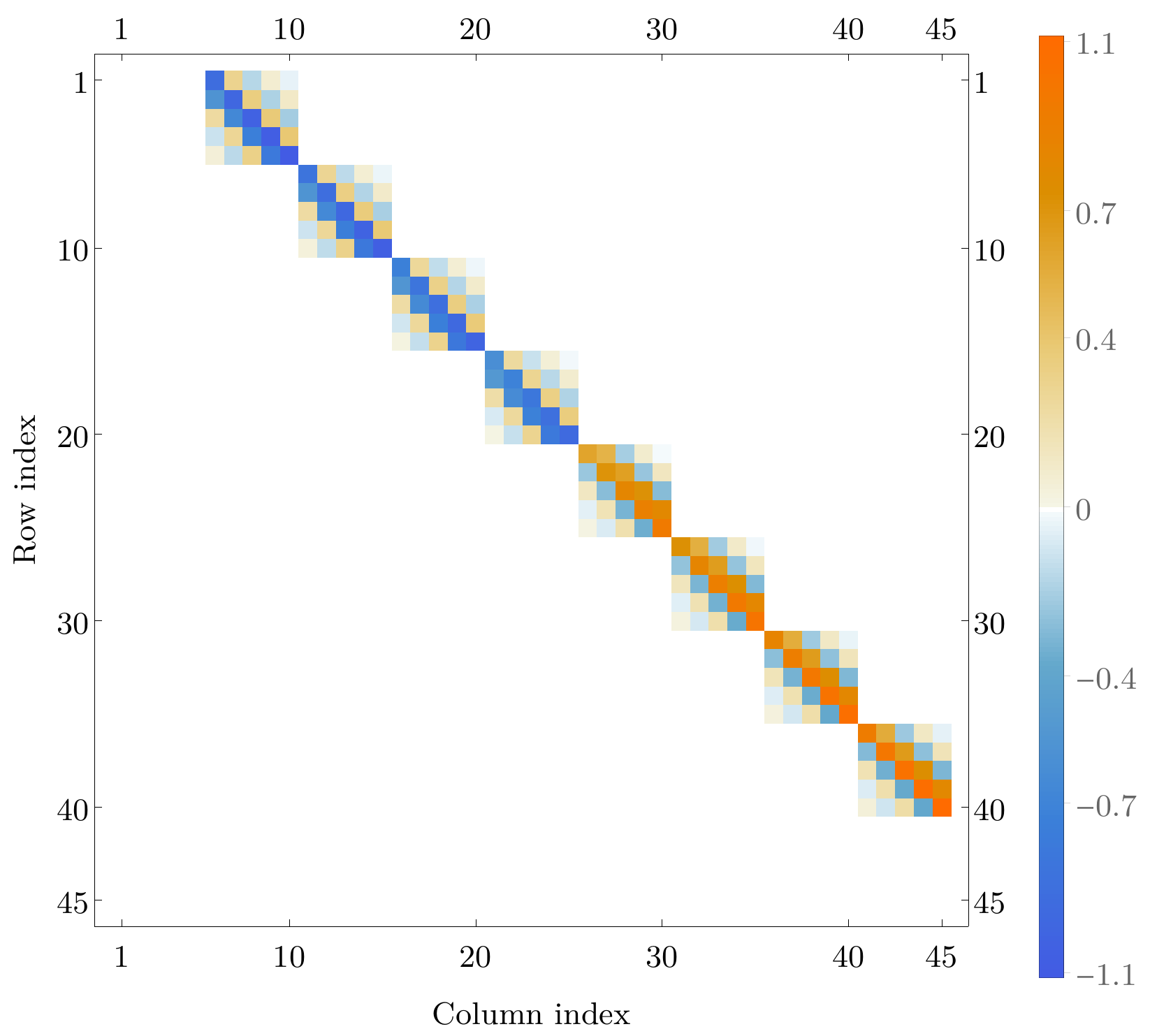}
\label{fig:PlotReL5a}}
\subfigure[~Imaginary part of $\mathbf{L}_1$]{
\includegraphics[width=0.475\columnwidth]{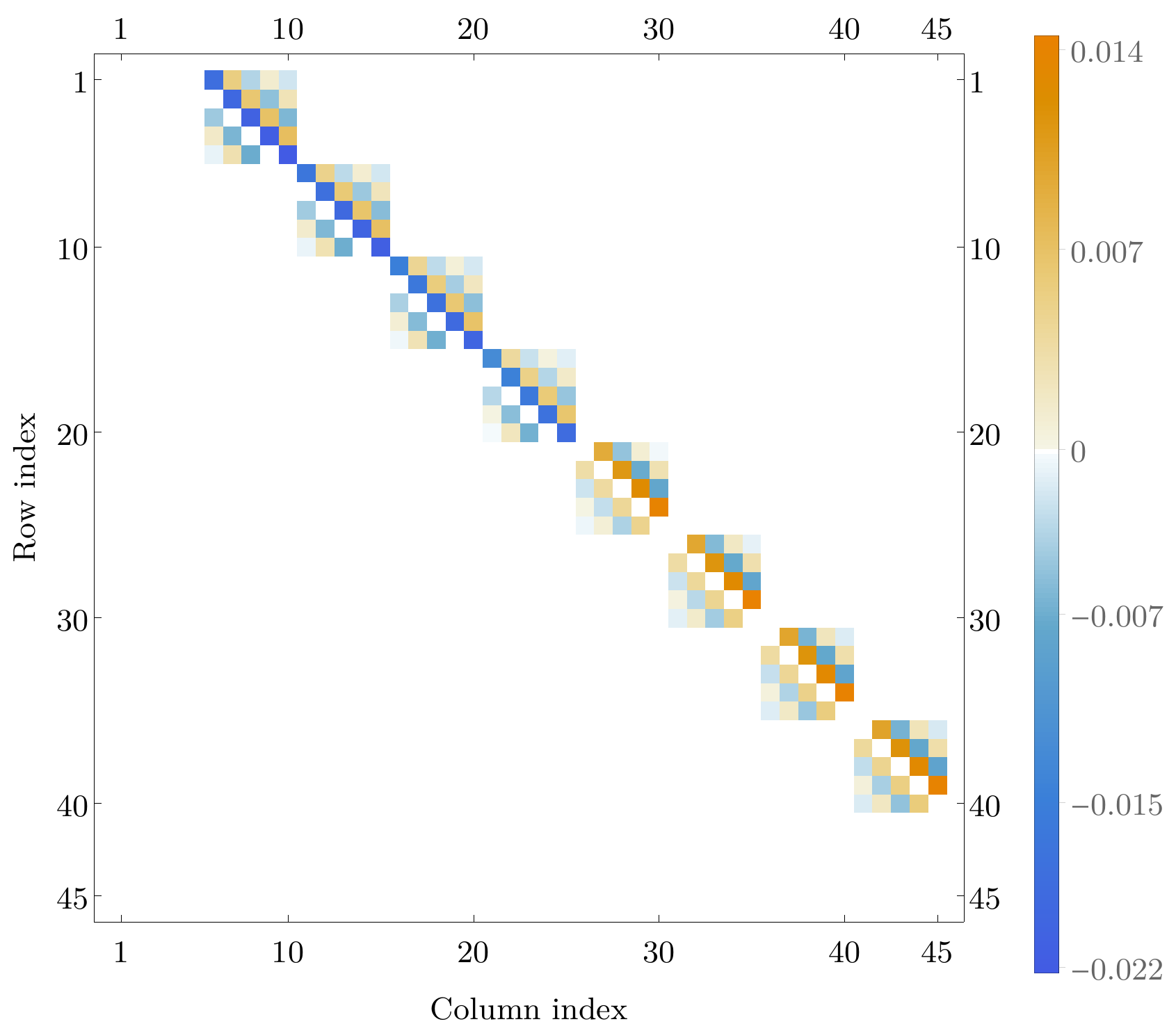}
\label{fig:PlotImL5a}}
\caption{Example of the Lindblad operator $\mathbf{L}_1$ when $\lambda=1.0\times10^{-6}$ m, $\omega_0=0.01$ m, $C_n^2=1.0\times 10 ^{-14}\;m^{-2/3}$, $t_z=100$, and $L=4$.}
\label{fig:PlotL5a}
\end{figure}

\begin{figure}[htbp]
\subfigure[~Real part of $\mathbf{L}_2$]{
\includegraphics[width=0.475\columnwidth]{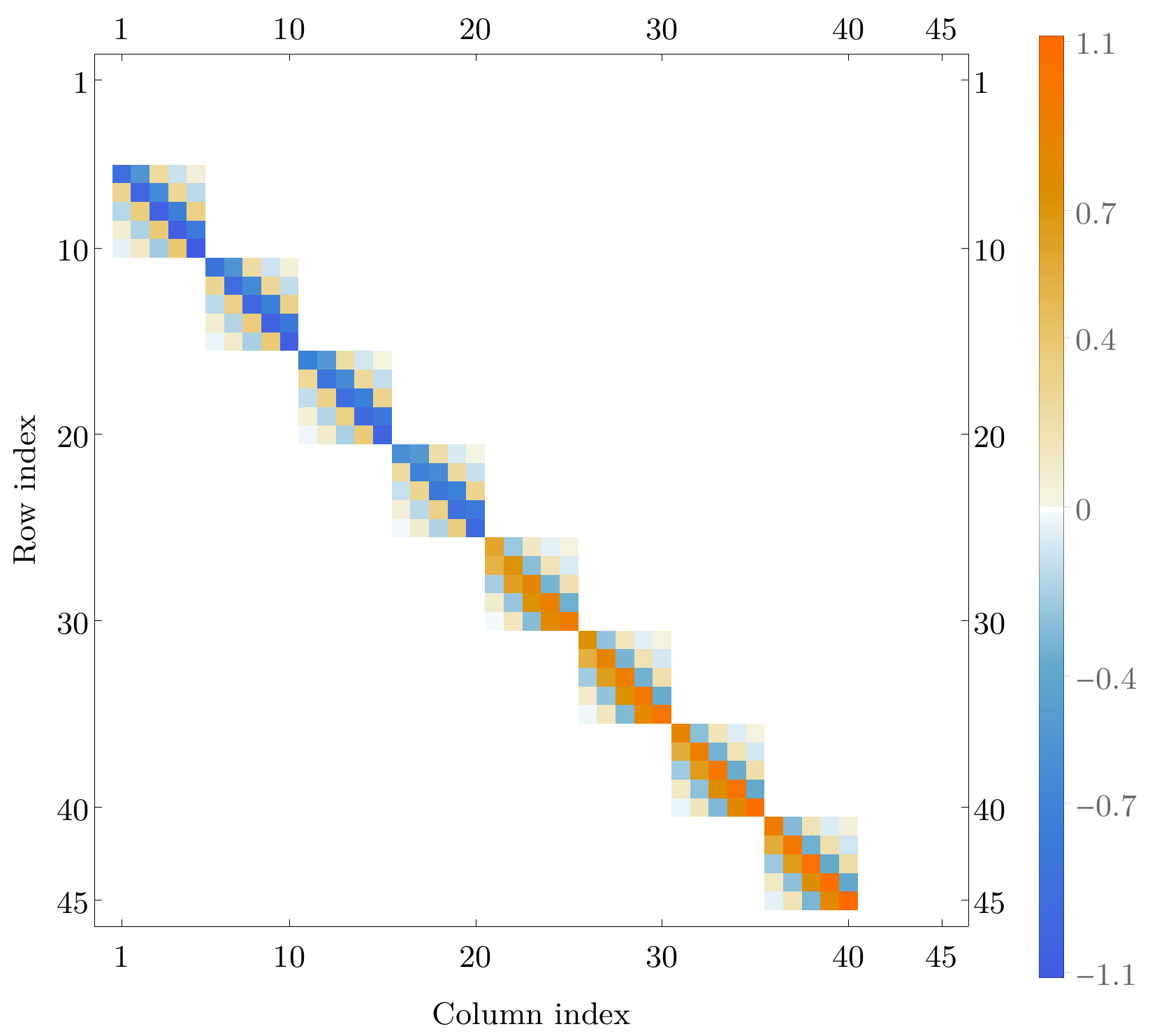}
\label{fig:PlotReL5b}}
\subfigure[~Imaginary part of $\mathbf{L}_2$]{
\includegraphics[width=0.475\columnwidth]{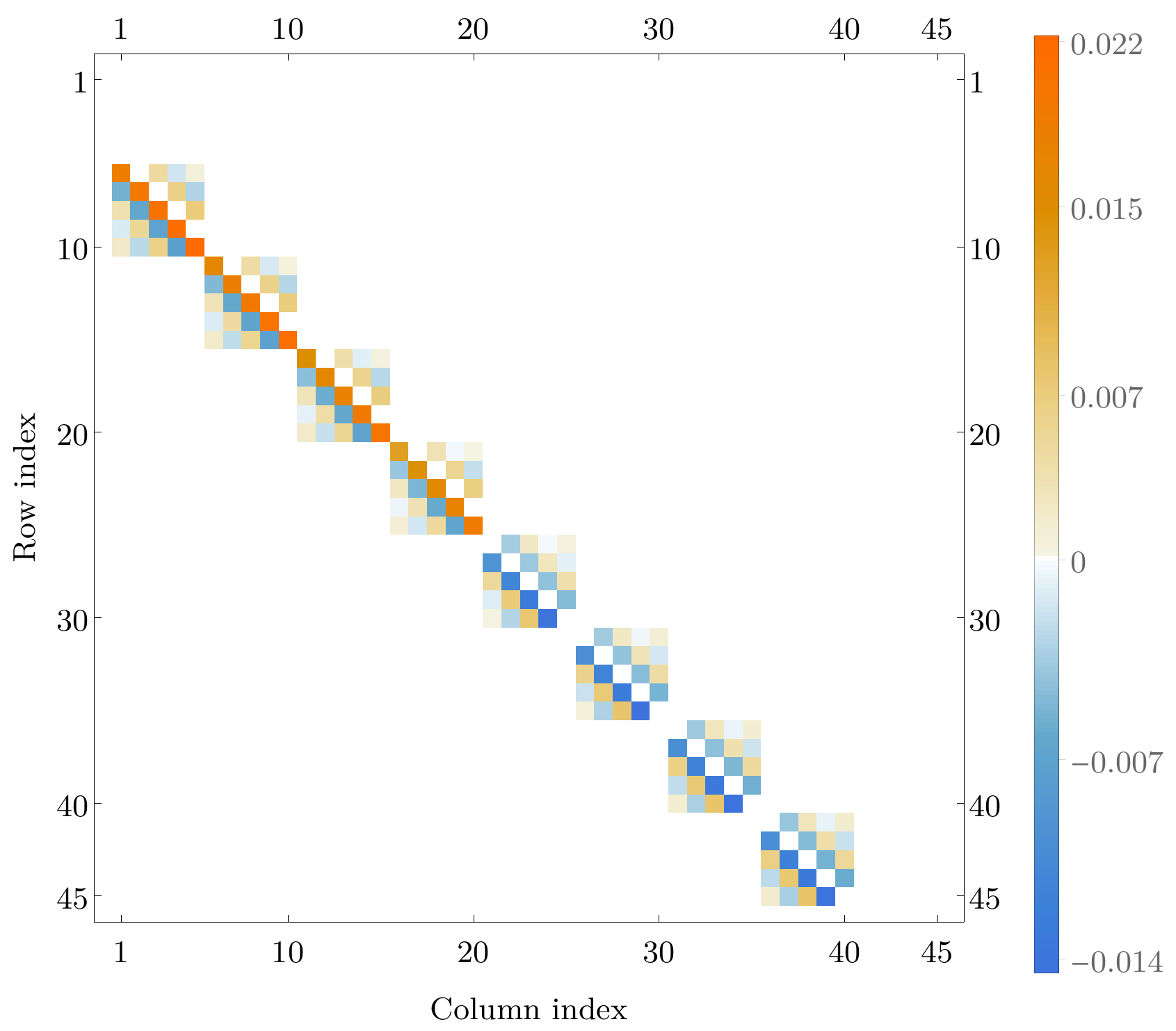}
\label{fig:PlotImL5b}}
\caption{Example of the Lindblad operator $\mathbf{L}_2$ when $\lambda=1.0\times10^{-6}$ m, $\omega_0=0.01$ m, $C_n^2=1.0\times 10 ^{-14}\;m^{-2/3}$, $t_z=100$, and $L=4$.}
\label{fig:PlotL5b}
\end{figure}

Additionally, we can appreciate in Fig. \ref{fig:EigvalMyProj5} that in the spectrum of $\tilde{D}$ the eigenvalues come in pairs. Each pair corresponds to a set of operators that raises or lowers the initial values of $l_m$ and $l_n$ by a set amount, and additionally changes the initial values of $r_m$ and $r_n$. Moreover, since the spectrum is dominated by the first two eigenvalues, this means that the first two Lindblad operators have an outsize influence on the noise process represented by the IPE.

\begin{figure}[htbp]
\subfigure[~Real part of $\mathbf{\tilde{L}}_1$]{
\includegraphics[width=0.475\columnwidth]{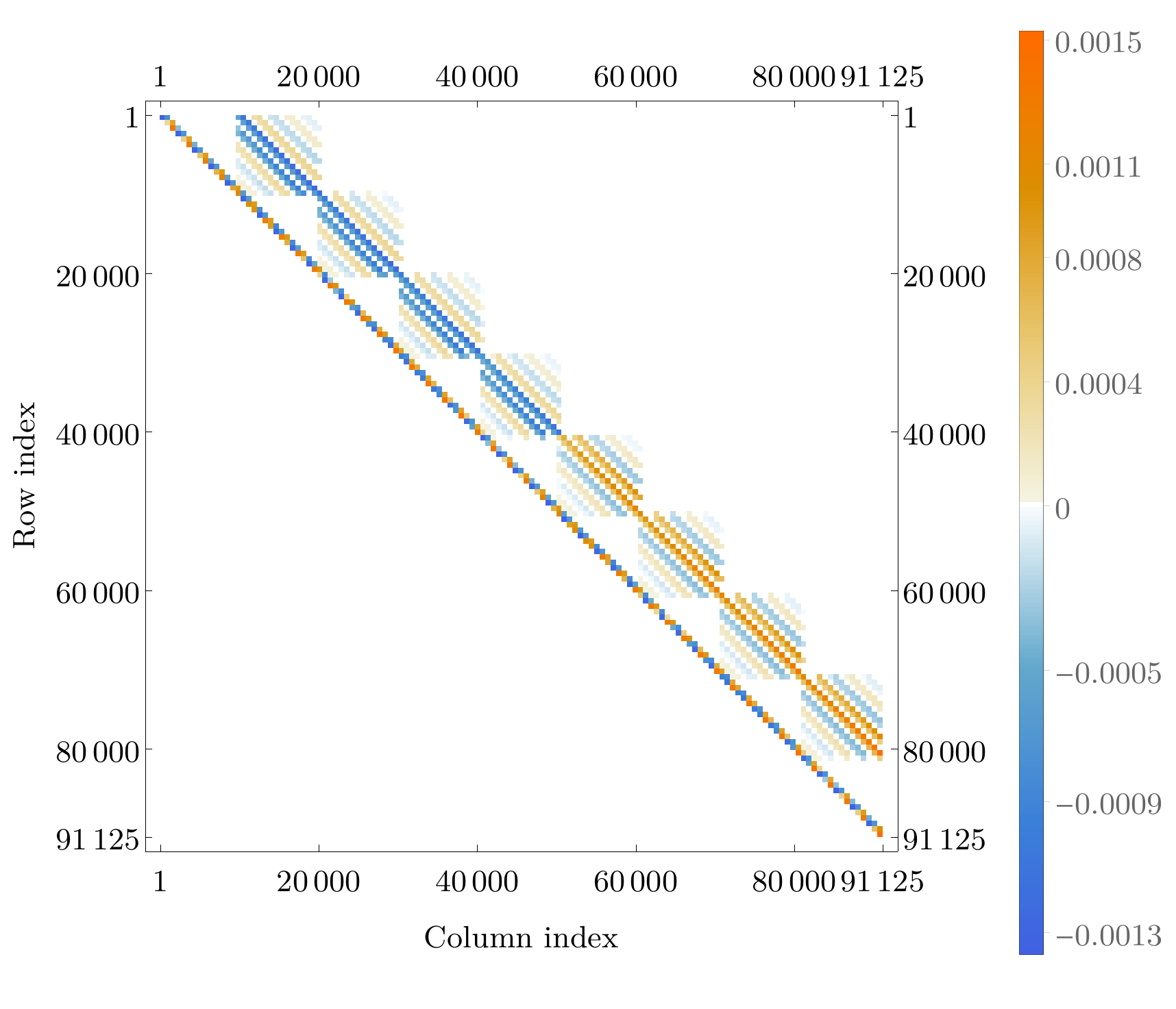}
\label{fig:PlotReLMultiple5a}}
\subfigure[~Imaginary part of $\mathbf{\tilde{L}}_1$]{
\includegraphics[width=0.475\columnwidth]{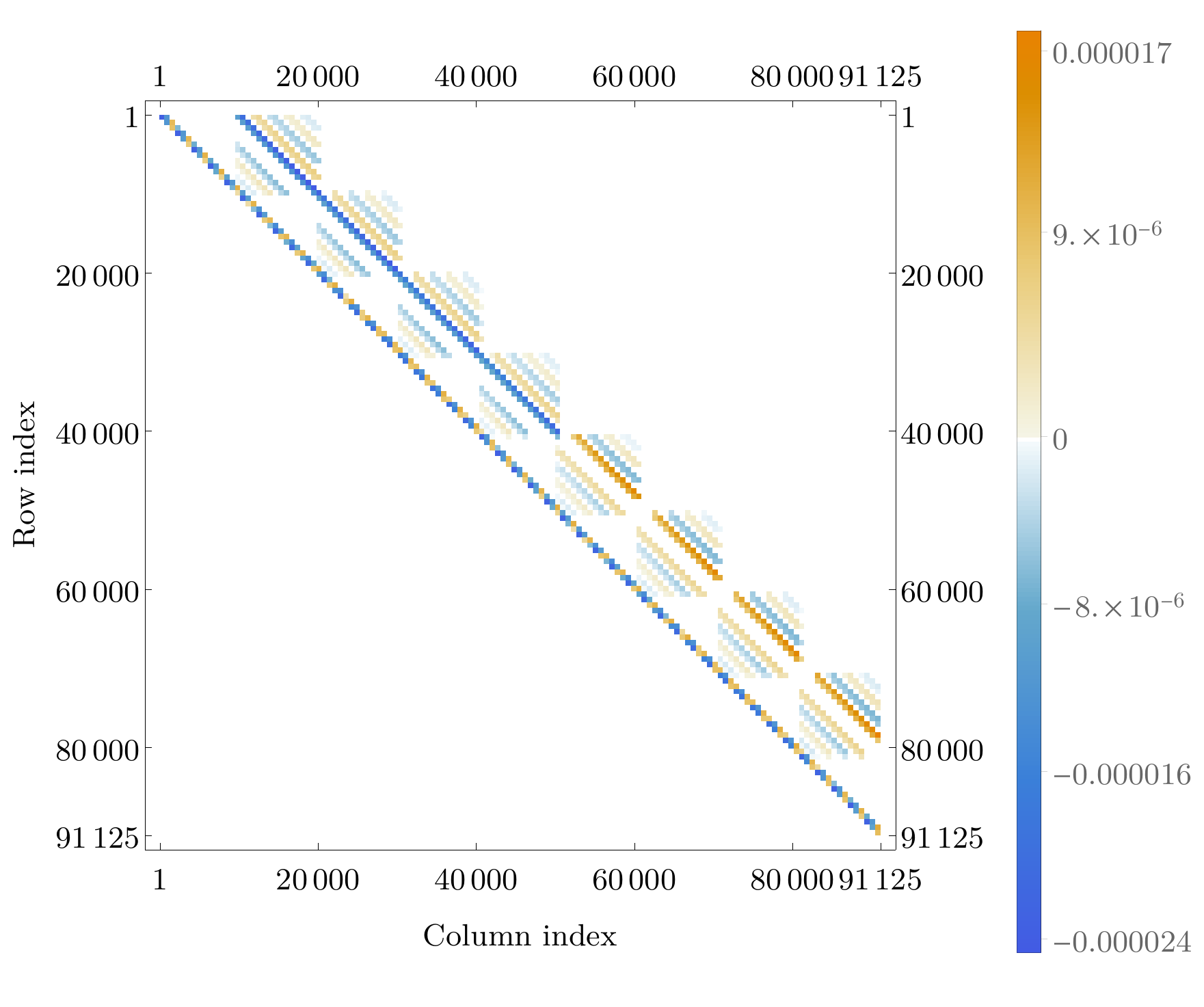}
\label{fig:PlotImLMultiple5a}}
\caption{Example of the Lindblad operator $\mathbf{\tilde{L}}_1$ when $\lambda=1.0\times10^{-6}$ m, $\omega_0=0.01$ m, $C_n^2=1.0\times 10 ^{-14}\;m^{-2/3}$, $t_z=100$, $L=4$, and $n=3$.}
\label{fig:PlotLMultiple5a}
\end{figure}

\begin{figure}[htbp]
\subfigure[~Real part of $\mathbf{\tilde{L}}_2$]{
\includegraphics[width=0.475\columnwidth]{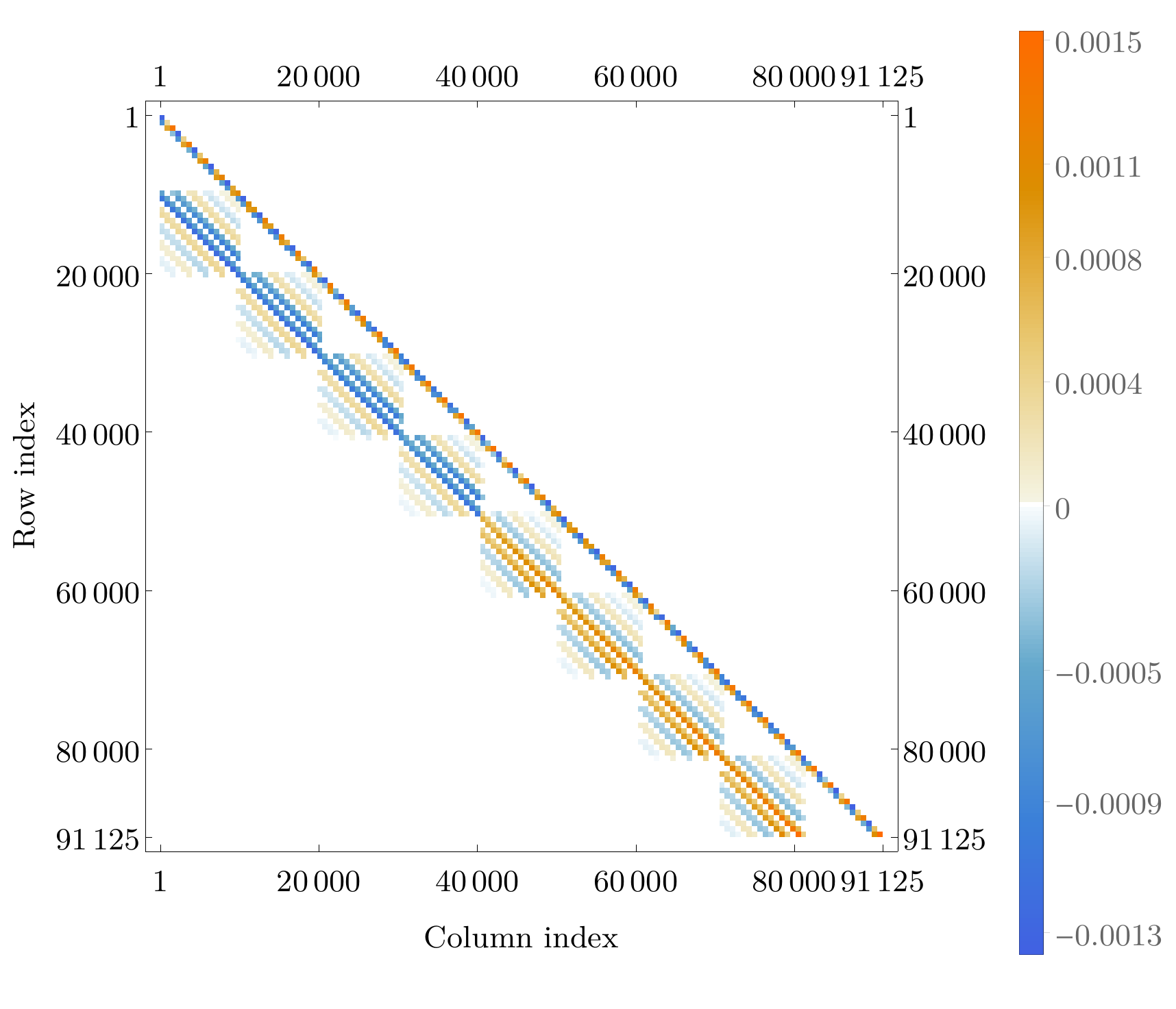}
\label{fig:PlotReLMultiple5b}}
\subfigure[~Imaginary part of $\mathbf{\tilde{L}}_2$]{
\includegraphics[width=0.475\columnwidth]{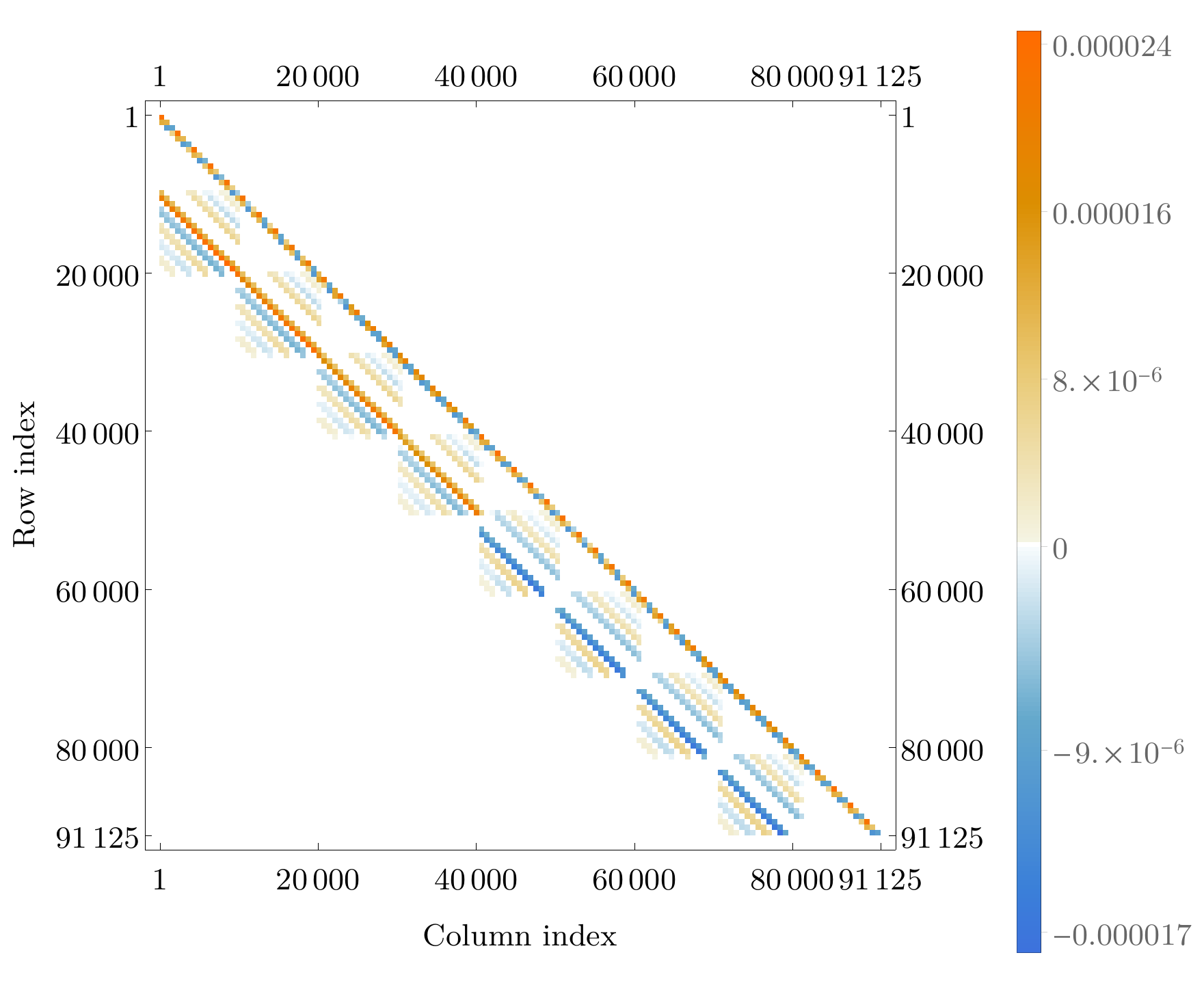}
\label{fig:PlotImLMultiple5b}}
\caption{Example of the Lindblad operator $\mathbf{\tilde{L}}_2$ when $\lambda=1.0\times10^{-6}$ m, $\omega_0=0.01$ m, $C_n^2=1.0\times 10 ^{-14}\;m^{-2/3}$, $t_z=100$, $L=4$, and $n=3$.}
\label{fig:PlotLMultiple5b}
\end{figure}

The two most dominant Lindblad operators are sparse, and represent a shift in $l_m,l_n$ by one unit, accompanied by
shifts in $r_m,r_n$ that depend on the values of OAM.  This can be seen from the blocks above and below the diagonal 
in Figs.~\ref{fig:PlotL5a} and \ref{fig:PlotL5b}. We have called these two operators $\mathbf{L}_1$ (raises the 
initial value of OAM) and $\mathbf{L}_2$ (lowers the initial value of OAM). 

Interestingly for our analysis involving an error detecting code in Sec. \ref{subsec:error_detect}, we have also seen that 
\begin{align}\label{eq:proplind1}
\Braket{l_m + 1 ,0|\mathbf{L}_1|l_m,0} = \overline{\Braket{l_m ,0|\mathbf{L}_2|l_m + 1, 0}},
\end{align}
and that
\begin{subequations}\label{eq:proplind2}
\begin{align}
\abs{\Braket{l_m + 1 ,0|\mathbf{L}_1|l_m,0}} &= \abs{\Braket{-l_m + 1, 0|\mathbf{L}_1|-l_m, 0}}\\
&=\abs{\Braket{l_m,0|\mathbf{L}_2|l_m + 1,0}}\\ 
&= \abs{\Braket{-l_m, 0|\mathbf{L}_1|-l_m + 1, 0}}.
\end{align}
\end{subequations}

The properties mentioned above seem also to extend to the multiphoton case, as can be seen from Figs.~\ref{fig:PlotLMultiple5a} and \ref{fig:PlotLMultiple5b}.

\begin{figure}[htbp]
\includegraphics[width=0.475\columnwidth]{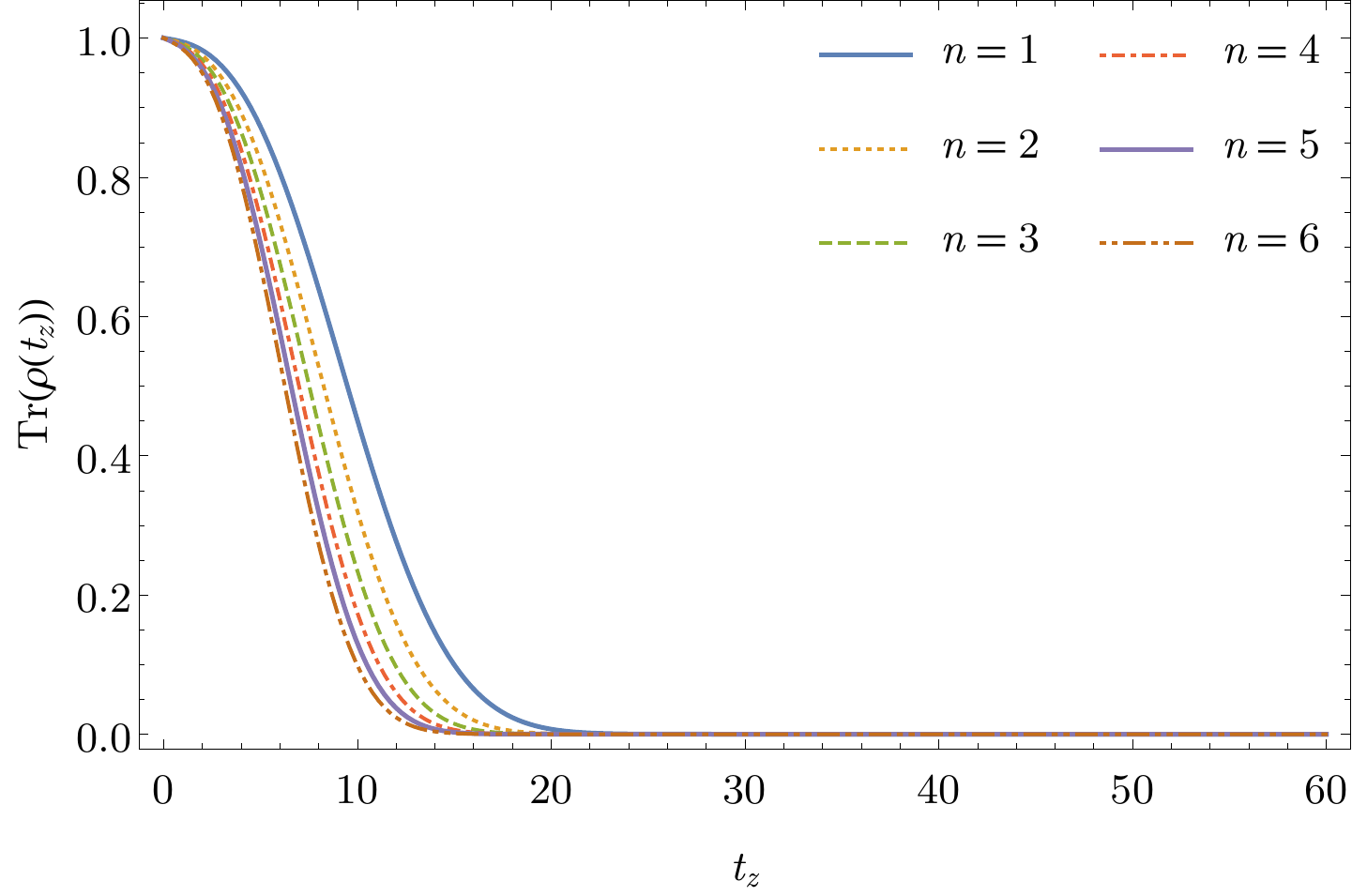}
\caption{Trace of the state in Eq. \eqref{eq:detecterror_state} after evolving for $t_z$ when $\lambda=1.0\times10^{-6}$ m, $\omega_0=0.01$ m, $C_n^2=1.0\times 10 ^{-14}\;m^{-2/3}$.}
\label{fig:MytrrftPlot}
\end{figure}

\section{Error Detection and Correction}

\subsection{An error-detecting code}\label{subsec:error_detect}

Let us now consider one-photon states for which the value of the radial  index is always zero.  We are going to use superposition of these states to build an error detecting code and investigate its performance.

\begin{figure}[htbp]
\includegraphics[width=0.475\columnwidth]{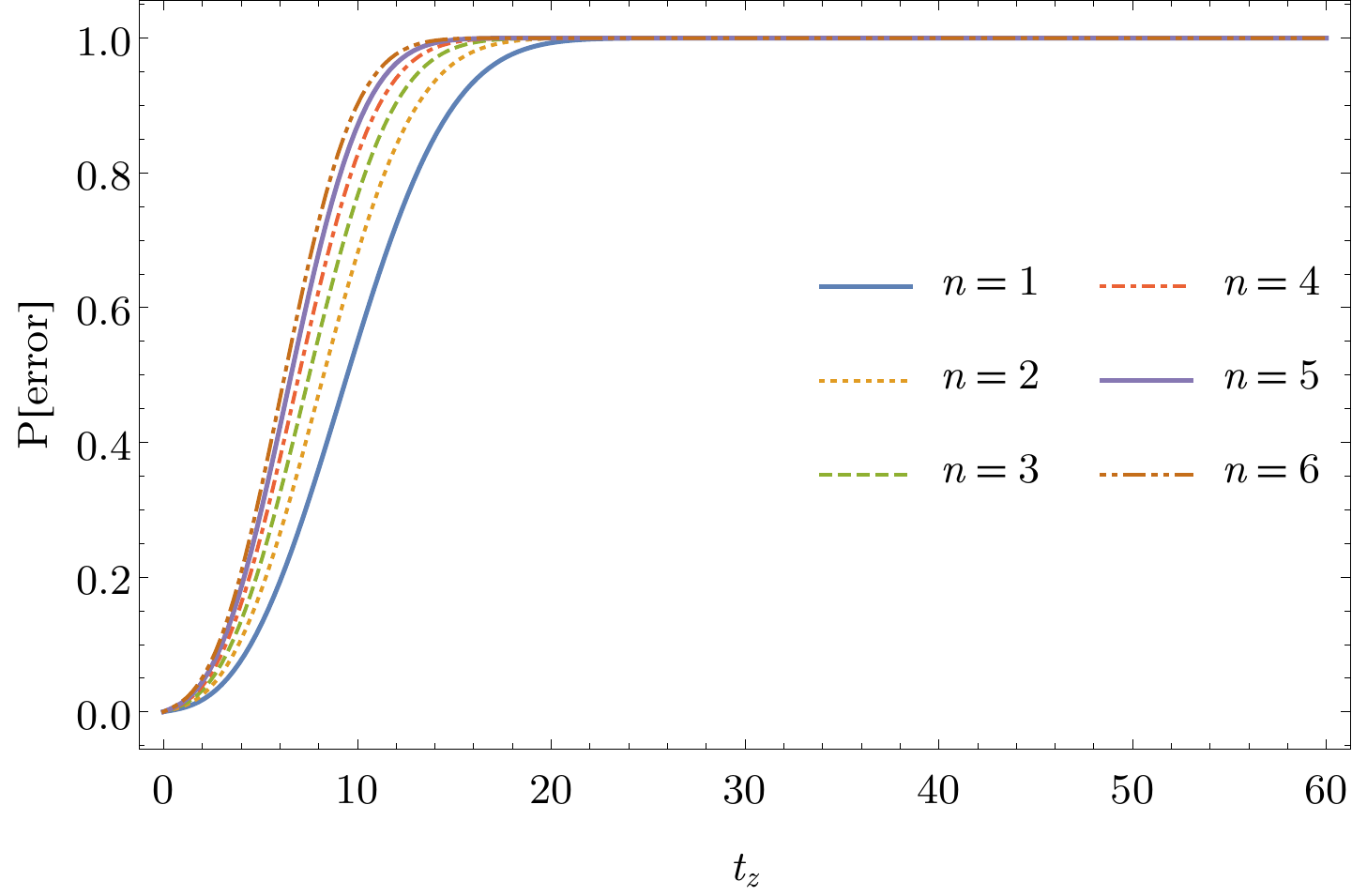}
\caption{Probability of a detectable error when using a state of the form given in Eq.~\eqref{eq:detecterror_state} after evolving for $t_z$ when $\lambda=1.0\times10^{-6}$ m,  $\omega_0=0.01$ m, $C_n^2=1.0\times 10 ^{-14}\;m^{-2/3}$.}
\label{fig:MyperrPlot}
\end{figure}

Consider the following states:
\begin{align}\label{eq:detecterror_state}
\ket{\psi_n} &= \alpha\ket{n,0} + \beta\ket{-n,0} 
\end{align}
where $\pm n$ are the azimuthal quantum numbers $l$ and $0$ is the radial quantum number $r$.  To illustrate the effects of the truncation of the state space, we can calculate the trace of the state $\rho(t_z)$ after evolving for a dimensionless propagation distance of $t_z=z\lambda/\pi\omega_0^2$, when the initial state is $\rho_n(0)=\ket{\psi_n}\bra{\psi_n}$. Since the Lindblad operators cause an initial state to scatter into neighboring modes, it is to be expected that, because of the truncation of the space, the trace of the final state will decrease with the propagation distance. We can see in Fig.~\ref{fig:MytrrftPlot} that this is precisely what happens. Moreover, we can also see that this effect depends on the choice of initial state. As the OAM of the initial state grows, the effect of the dissipation becomes larger.

\begin{figure}[htbp]
\includegraphics[width=0.475\columnwidth]{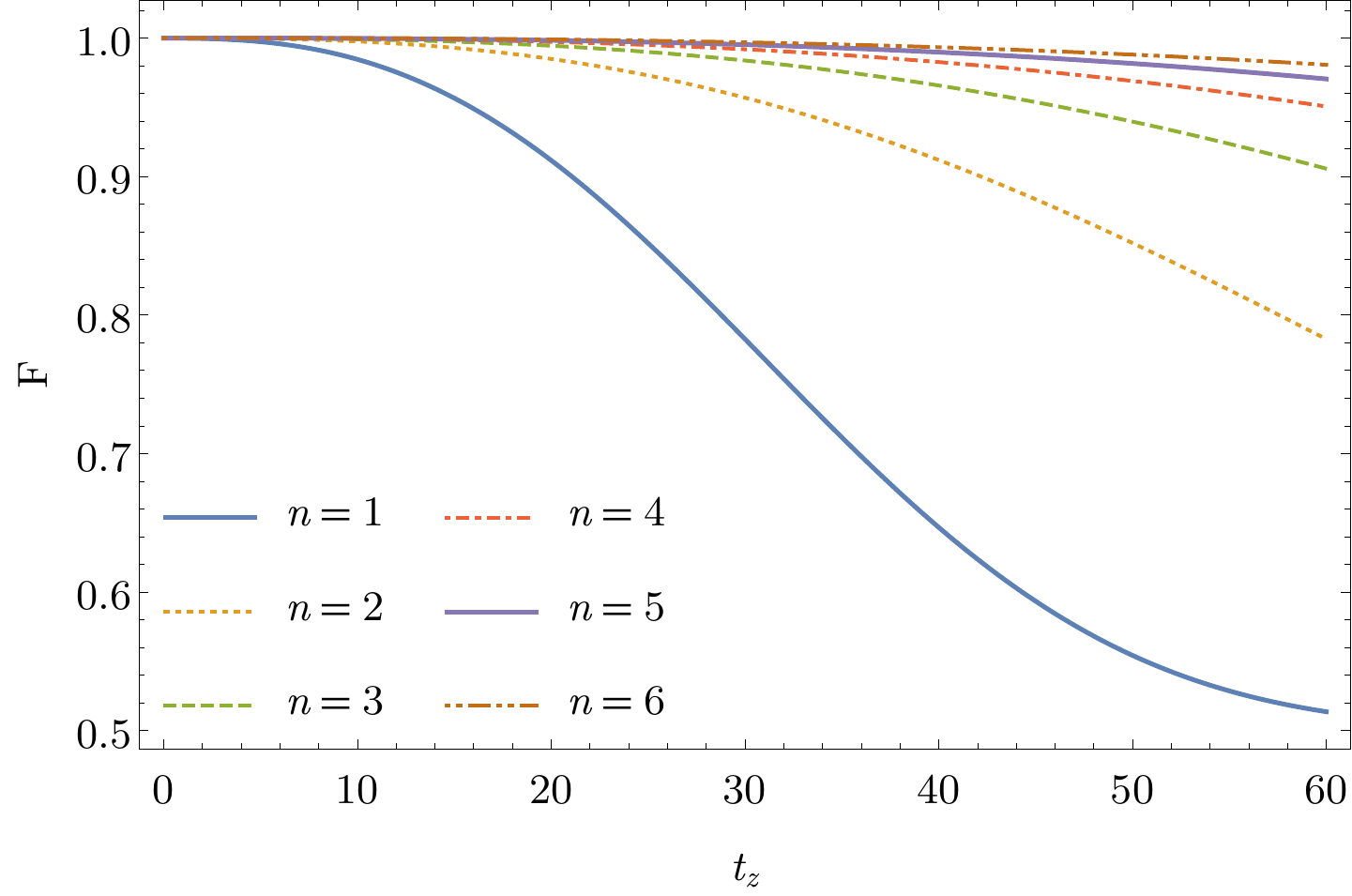}
\caption{Minimum fidelity between an initial state of the form given in Eq. \eqref{eq:detecterror_state}
and the state after evolving for $t_z$ when $\lambda=1.0\times10^{-6}$ m, $\omega_0=0.01$ m,
$C_n^2=1.0\times 10 ^{-14}\;m^{-2/3}$.}
\label{fig:MinFNDPlot}
\end{figure}

Since the most relevant Lindblad operators by magnitude are those that shift the OAM value by one unit, we can see 
that if we restrict our attention to these operators and use states of the form given by \eqref{eq:detecterror_state}
to build a code, then terms in the Lindblad equation that correspond to $\mathbb{I}$, $\mathbf{L}_1\mathbf{L}_2$, or 
$\mathbf{L}_2\mathbf{L}_1$ correspond to no error (as is clear from the properties in Eqs. \eqref{eq:proplind1}
and \eqref{eq:proplind2}), whereas terms that either change the initial value of $r_m,r_n$ or take us out of the 
truncated space represent a detectable error. We can numerically calculate the probability of such an error as a 
function of the propagation distance. The results of this calculation are seen in Fig. \eqref{fig:MyperrPlot}. As 
can be seen, as the propagation distance increases, so does the probability of detecting an error.

Of course, it is also possible for {\it undetectable} errors to occur.  However, the structure of the Lindblad operators (as described above) means that the dominant error processes shift the values of $l$, resulting in a detectable error.  They do not produce relative phase errors between states with azimuthal quantum numbers $\pm l$, and the amplitude of transitions out of the code space is the same for $\ket{n,0}$ and $\ket{-n,0}$.  So undetectable errors result mostly from higher-order terms in the evolution.  At short ranges, errors are dominated by terms that are first order in $\mathbf{L}_{1,2}$; but at longer ranges, higher-order errors become important.

We can assess the importance of undetectable errors by calculating the fidelity between the initial and a renomalized final state varies as a function of the propagation distance for different values of OAM of the initial state. For each point in Fig.~\ref{fig:MinFNDPlot} we chose the values of $\alpha$ and $\beta$ that minimize the fidelity.  It is clear and consistent with our expectations that as the propagation distance increases, the minimum fidelity decreases.  We also see that fidelity falls off more slowly as we increase the value of $n$, the azimuthal quantum number used in the error-detection code.

\subsection{A scheme for error correction in OAM of photons}

This construction of an error-detecting code for OAM suggests one possible approach to error correction.  This approach builds on standard QECCs by concatenating them with the type of error-detecting code described in the previous subsection.

The idea is quite simple.  An $[[n,k,d]]$ QECC encodes $k$ logical qubits into $n$ physical qubits and has minimum distance $d$ \cite{Lidar-Quantum-2013-0}. Such a code can correct general errors on up to $\left\lfloor{\frac{d-1}{2}}\right\rfloor$ qubits.  However, such codes can also correct up to $d-1$ {\it erasure errors}:  that is, errors in which $d-1$ qubits are erased (that is, completely randomized or lost), but where it is known which qubits have been erased.  This means that if one knows which qubits in a codeword have errors, but not necessarily what the errors are, then one could discard those qubits, replace them with new qubits in any state, and then carry out the correction procedure for erasures.  Knowing which qubits have errors makes a code more powerful, able to correct twice as many errors.

A natural scheme presents itself.  The physical qubits of the $[[n,k,d]]$ are realized by $n$ closely-spaced OAM photons using the quantum error-detection code from the previous subsection.  When these photons are received, one first measures to detect whether an error has occurred on each of the $n$ photons.  If an error is detected, that photon can be discarded and replaced by another photon in an arbitrary state in the code space of the error-detecting code.  Provided that no more than $d-1$ photons have errors, they can be corrected as erasure errors.

Let's see how we can model these erasure errors, treating the photons as qubits:  $\ket0 \equiv \ket{n,0}$, $\ket1 \equiv \ket{-n,0}$.  Suppose an error is detected on a given photon; we discard that photon and replace it with the maximally mixed state $I/2 = (1/2)(\ket0\bra0 + \ket1\bra1)$.  This process (discarding the erroneous photon and replacing it with a maximally mixed state) can be modeled by a completely depolarizing map:
\begin{equation}
\rho \rightarrow \frac{1}{4} \left( \rho + X\rho X + Y\rho Y + Z\rho Z \right) = \frac{I}{2},
\end{equation}
where $X$, $Y$, $Z$ are the usual Pauli matrices.  So we can treat an erased qubit as having been affected by an $X$, $Y$, $Z$, or no error with equal probability.  The key is that we know {\it which} qubit (or qubits) may have an error, which means that the given error-correcting code can correct more errors than if they act on unknown qubits.

Let's look at a specific example to see how this works.  Suppose we concatenate the error-detecting code described above for $n=\pm1$ with the $[[7,1,3]]$ Steane code \cite{Steane-Error-1996-0}.  Normally, the Steane code can correct an arbitrary error on a single qubit.  This is equivalent to being able to correct the error set $\{ I, X_1, Y_1, Z_1, \ldots, X_7, Y_7, Z_7 \}$.  Here, $X_j$ is the Pauli $X$ acting on qubit $j$, and so forth.  An arbitrary error $E_j$ on qubit $j$ can be written as a linear combination $E_j = a I + b X_j + c Y_j + d Z_j$.

The Steane code is an example of a Calderbank-Shor-Steane (CSS) code \cite{Calderbank-Good-1996-0,Steane-Simple-1996-0}.  Essentially, it works by combining two classical linear codes.  The code space of a CSS code is the intersection of a code that corrects bit flips ($X$ errors) and a code that corrects phase flips ($Z$ errors); a $Y$ error is a combination of a bit flip and a phase flip on the same qubit (up to a global phase), $Y=iXZ$.  In the case of the Steane code, both of these codes are versions of the 7-bit Hamming code, whose binary parity-check matrix is
\begin{equation}
\mathbf{H} = \left(\begin{array}{ccccccc} 1 & 0 & 0 & 0 & 1 & 1 & 1 \\
0 & 1 & 0 & 1 & 0 & 1 & 1 \\ 0 & 0 & 1 & 1 & 1 & 0 & 1 \end{array} \right) .
\end{equation}
An easy way to see that this code can correct a single error on any qubit is to note that the columns of $\mathbf{H}$ are all distinct.  These columns are the {\it error syndromes} that will be measured when the corresponding bit is flipped.  So the Steane code can detect and correct a single bit flip {\it and} a single phase flip on any qubit, and hence can correct the error set listed above (and indeed, some additional errors as well, such as an $X$ error on one qubit and a $Z$ error on another).  However, it cannot correct a general error on two qubits.  For instance, suppose bit flips occur on qubits 1 and 2.  The error syndrome will be the linear combination of the first two columns of $\mathbf{H}$, which is the same as column 6.  So the code cannot distinguish this weight-2 error from a different weight-1 error.

What if instead qubits 1 and 2 are erased?  In this case, we {\it know} which two qubits are affected.  Treating the erasures as depolarizations, the set of possible errors becomes $\{ I, X_1, Y_1, Z_1, X_2, Y_2, Z_2, X_1 X_2, X_1 Y_2, X_1 Z_2, \ldots, Z_1 Z_2 \}$.  All 16 errors in this set have distinct error syndromes, and hence they can all be corrected.  We can see this again by looking at the parity check matrix $\mathbf{H}$.  The bit flips and phase flips are again corrected separately.  Erasure of qubits 1 and 2 could produce $X$ errors on qubit 1, on qubit 2, on both qubits, or neither.  Therefore the error syndromes will be the 4 linear combinations of the first 2 columns, which are all distinct.  There can be no confusion between errors on qubits 1 and 2 and an error on qubit 6, because we know that the erasures acted on qubits 1 and 2 and not on 6.

How would this scheme perform with the noise process derived in this paper?  At moderate ranges, the error process is dominated by the first two Lindblad operators $\mathbf{L}_{1,2}$.  We can consider a perturbation expansion of the evolution
\[
\rho(t) = \exp(\mathbb{L}t) \rho(0) ,
\]
where $\mathbb{L}=\mathbb{C} + \mathbb{D}$ is the Liouvillian derived earlier this paper.  Assume this noise now acts independently on all 7 photons of the codeword.  The first order in perturbation theory produces only detectable errors on a single photon; these correspond to single erasures, which can be corrected.  Second order in perturbation theory can produce detectable errors on two qubits, which can also be corrected (as two erasures).  At second order, we can also produce {\it undetectable} errors, resulting from a pair of errors on the same photon.  However, these errors can {\it also} be corrected:  if no errors are detected, then the Steane code can be used in the usual way as an error-correcting code that can an arbitrary error on a single qubit.  So we see this scheme can correct the error process up to second order in perturbation theory.

It is quite possible to extend this scheme in a number of ways.  One could use another error-correcting code, that encodes more qubits and/or corrects more errors.  Concatenating an $[[n,k,d]]$ quantum error correcting code with the error-detecting code will generically double the number of errors that can be corrected.  Since the OAM space is infinite-dimensional (in principle), one can also encode qudits rather than qubits, and use a suitable qudit code concatenated with a qudit quantum error-detecting code.  It is also possible to store a qubit in the polarization of each photon; these qubits are insensitive to the effects of turbulence, and as long as the photons are not lost they do not require correction.

It is quite possible that much better QECCs than this can be designed for OAM of photons.  We have not made significant use of the permutation symmetry of the noise for closely-spaced photons, which intuitively should be possible to exploit for better performance.  But the straightforward approach described in this section should work reasonably well if the noise is not too strong.

\section{Discussion and Conclusions} 

We have seen that for the infinitesimal propagation equation derived in \cite{Roux-Infinitesimal-propagation-2011-0,Roux-Erratum:-2013-0} it is possible to use the techniques from \cite{Havel-Robust-2003-0} to obtain a discrete Lindblad equation for the effects of Kolmogorov turbulence on an OAM photon propagating through a turbulent atmosphere. A numerical analysis of the Lindblad operators reveal that they come in pairs, and represent shifts in the OAM content of a state. The dominant pair (as measured by the magnitude of the corresponding eigenvalue) shifts the initial $l_m,l_n$ by one and also can change the values of $r_m,r_n$.

Based on this analysis of the dominant errors, we presented a simple quantum error-detecting code, and showed how this could be concatenated with an $[[n,k,d]]$ QECC to given an error-correction procedure for quantum information encoded across the OAM of multiple photons.

We also found the form of the Lindblad operators when there are multiple photons propagating with a time separation that is less than the characteristic time of the turbulence process. Interestingly, these Lindblad operators---which can be interpreted as distinct error processes---act collectively on the $n$ photons in a permutation-symmetric manner.

This raises an intriguing question in quantum error correction.  Given the permutation symmetry of the photon errors, is it possible to build a quantum error-correcting code for information encoded across multiple photons that exploits the symmetry of the noise?  For finite-dimensional spaces such permutation symmetry might be expected to give rise to a decoherence-free subspace or noiseless subsystem \cite{Zanardi-Noiseless-1997-0,Lidar-Decoherence-Free-1998-0,Lidar-Review-2012-0}.  In an infinite-dimensional space---such as occurs in OAM---it seems that this need not be true.  But it seems likely that the symmetry could still be exploited by a properly designed QECC.  While we have not yet found such a scheme, we believe this is a promising topic for further research.

Another intriguing possibility is to combine quantum error correction with a method such as adaptive optics to achieve better performance in the face of turbulence \cite{Gonzalez-Alonso-Recovering-2016-0}.  Indeed, given the high noise rates from atmospheric turbulence \cite{Gonzalez-Alonso-Protecting-2013-0}, it may be necessary to combine adaptive optics and quantum error-correction to allow practical quantum communication at all.  These topics are also the subject of ongoing research.

\section{Acknowledgements}

This research was supported by the ARO MURI under Grant No. W911NF-11-1-0268 
and by NSF Grant No. CCF-1421078.

\bibliographystyle{apsrev4-1}
\bibliography{./References/Reference_Database}

\end{document}